\documentclass[useAMS]{mn2e}
\usepackage{graphics}
\usepackage{graphicx}
\usepackage{epsf}
\usepackage{url}
\newif\ifAMStwofonts
\AMStwofontstrue

\newcommand{\lines}{
  \begin{figure}
    \begin{center}
      \leavevmode
      \epsfxsize=\columnwidth    
      \epsfbox{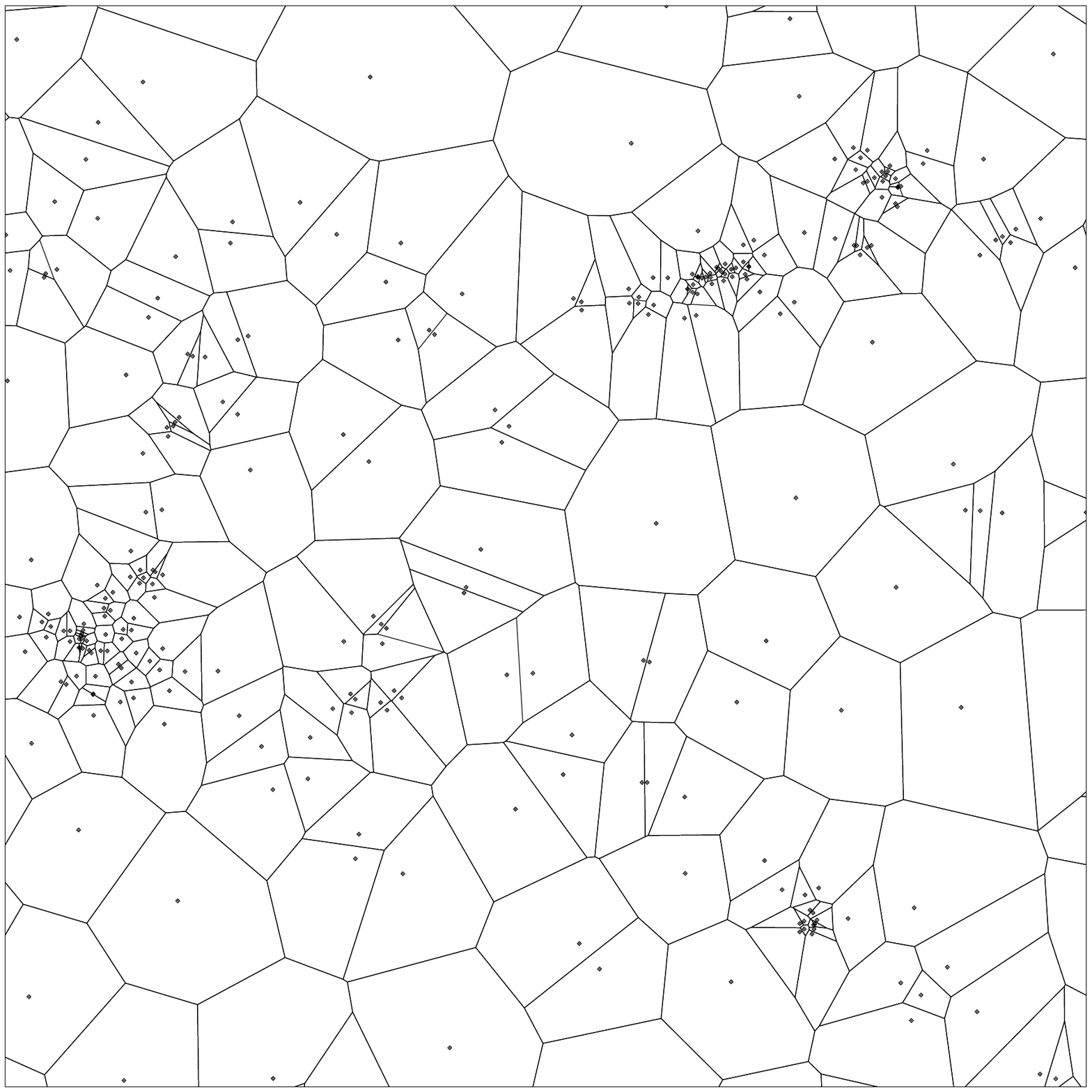}
    \end{center}
    \caption[1]{ \small A two-dimensional Voronoi diagram of particles from
    an $N$-body simulation.
    \label{lines}
    }
  \end{figure}
}

\newcommand{\guard}{
  \begin{figure}
    \begin{center}
      \leavevmode
      \epsfxsize=\columnwidth    
      \epsfbox{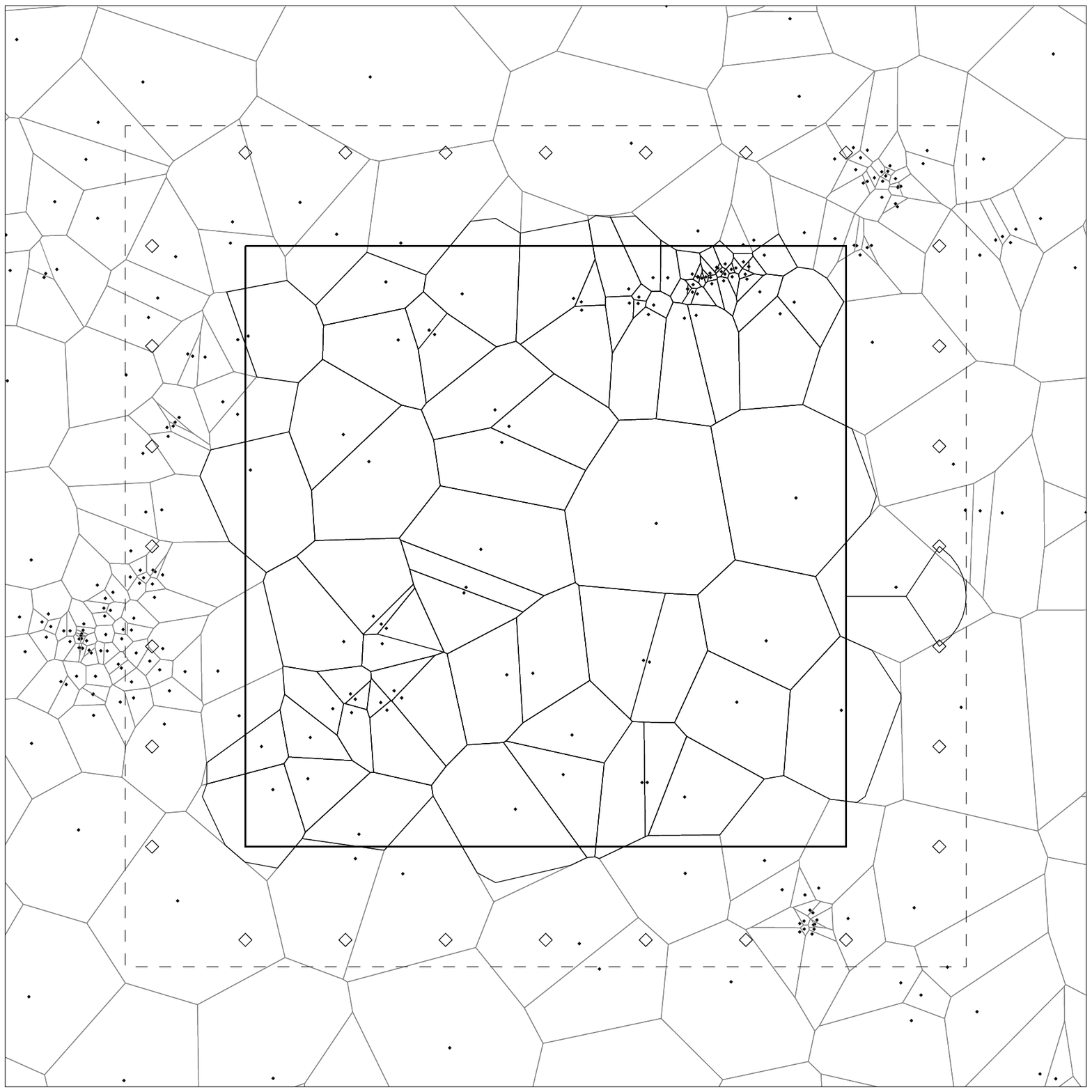}
    \end{center}
    \caption[1]{ \small The same region as in Figure \ref{lines}, with a
    Voronoi diagram on a sub-box superimposed.  Solid lines demarcate the
    sub-box, with the buffer around it outlined by dashed lines.  Guard
    points in the buffer appear as diamonds.  The Voronoi cells calculated
    using all particles appear in grey, while the Voronoi cells of
    particles in the sub-box calculated using only the sub-box, buffer, and
    guard particles appear in black.  There are a few discrepancies,
    notably at the center on top, and on both sides near the bottom.
    Discrepancies indicate that the buffer should be enlarged, and the VD
    recalculated.
    
    \label{guard}
    }
  \end{figure}
}
\newcommand{\guardclose}{
  \begin{figure}
    \begin{center}
      \leavevmode
      \epsfxsize=2in
      \epsfbox{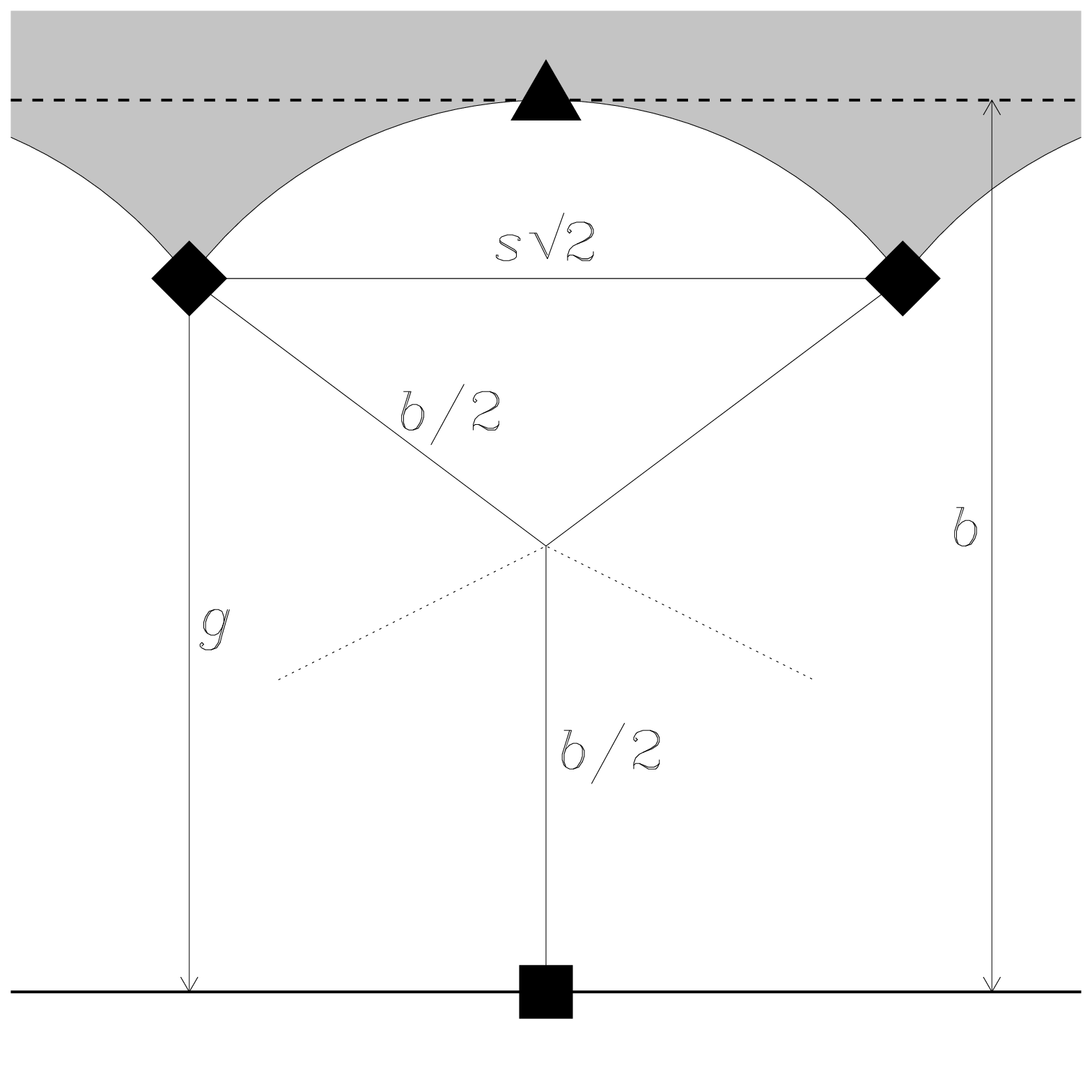}
    \end{center}
    \caption[1]{ \small A diagram showing the calculation of $g$, the
    optimal distance from the sub-box (at the bottom of the diagram) to the
    guard points.  The guard points preempt alteration of the sub-box
    Voronoi diagram, ensuring that points outside the buffer, of width $b$,
    cannot affect it.  The diamonds are guard points diagonally adjacent on
    a two-dimensional grid of spacing $s$, offset by a distance $g$ from a
    face of the sub-box.  The hardest scenario for the guard points to
    preempt has a particle in the sub-box at the square, and a particle
    outside the buffer at the triangle, which would have been one of the
    square's neighbors if it were inside the buffer.  The guard points
    preempt any potential neighbors to the square which lie in the shaded
    region.  The dotted lines are perpendicular bisectors between the
    diamonds and the square.

    \label{guardclose}
    }
  \end{figure}
}

\newcommand{\tank}{
  \begin{figure}
    \begin{center}
      \leavevmode
      \epsfxsize=\columnwidth    
      \epsfbox{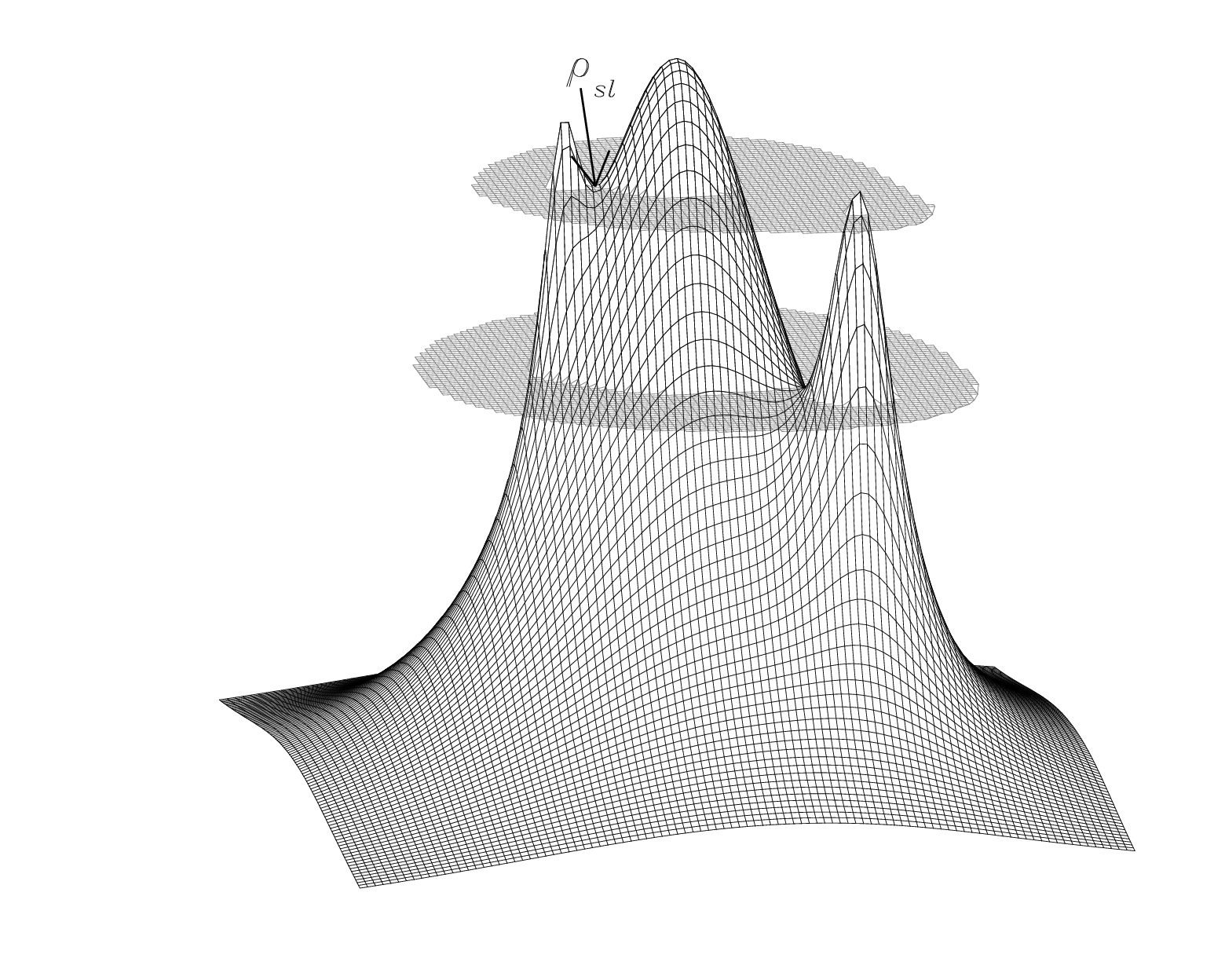}
    \end{center}
    \caption[1]{ \small A schematic density field containing a halo with
    two subhaloes.  If we put this surface in a water tank, and gradually
    reduce the water level, the central peak will be the first to emerge.
    As we reduce the water level, no higher peak is uncovered, so the
    entire region belongs to the central peak.  If we reduce the water
    level from the peak of the left subhalo, we will reveal a landbridge to
    a higher peak when the water reaches the upper grey plane.  If we do
    the same for the subhalo on the right, a landbridge to a higher peak
    will appear at the lower grey plane.  We define the boundaries of each
    subhalo to be a density contour at the lowest density on the landbridge
    (at the height of its grey plane), called the strongest link density,
    $\rho_{sl}$.  The probability that a subhalo is real depends on the
    ratio of its peak density to $\rho_{sl}$.

    \label{tank}
    }
  \end{figure}
}  

\newcommand{\zoning}{
  \begin{figure}
    \begin{center}
      \leavevmode
      \epsfxsize=2in    
      \epsfbox{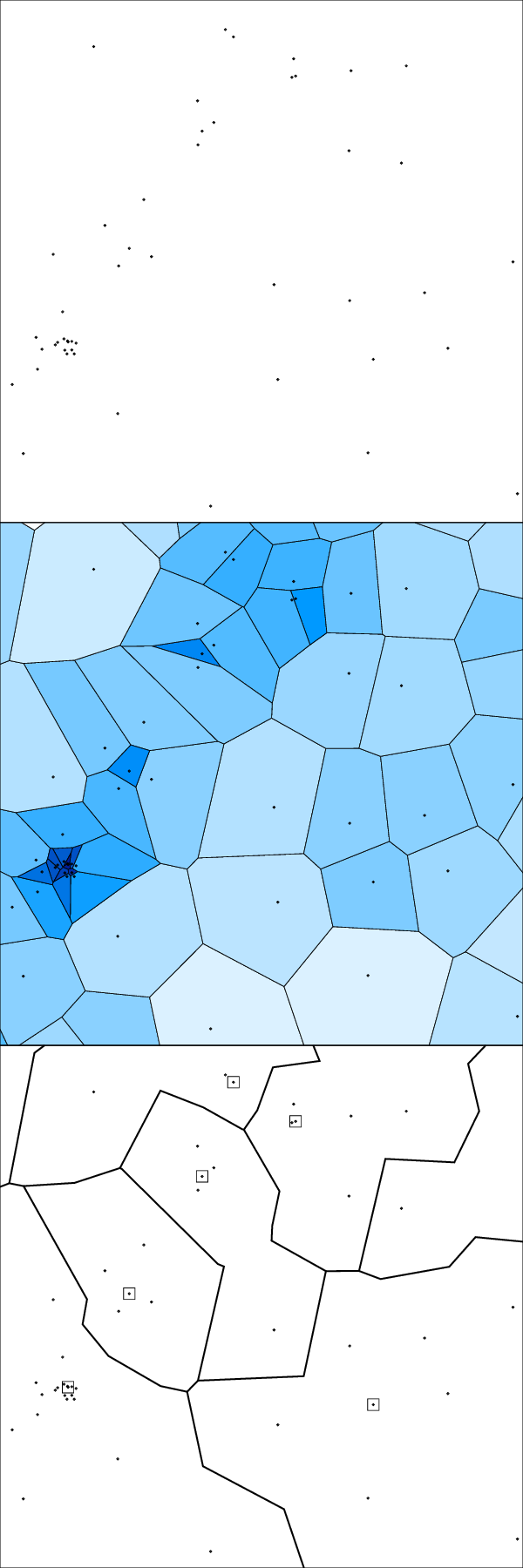}
    \end{center}
    \caption[1]{ \small Zoning.  The top panel shows a raw set of
    particles.  The middle panel shows the 2-D Voronoi diagram of these
    particles, with cells shaded according to their areas.  The bottom
    panel shows how these particles are partitioned into zones, with the
    peak of each zone indicated with a square.
      \label{zoning}
    }
  \end{figure}
}
  \newcommand{\pcdf}{
  \begin{figure}
    \begin{center}
      \leavevmode
      \epsfxsize=\columnwidth    
      \epsfbox{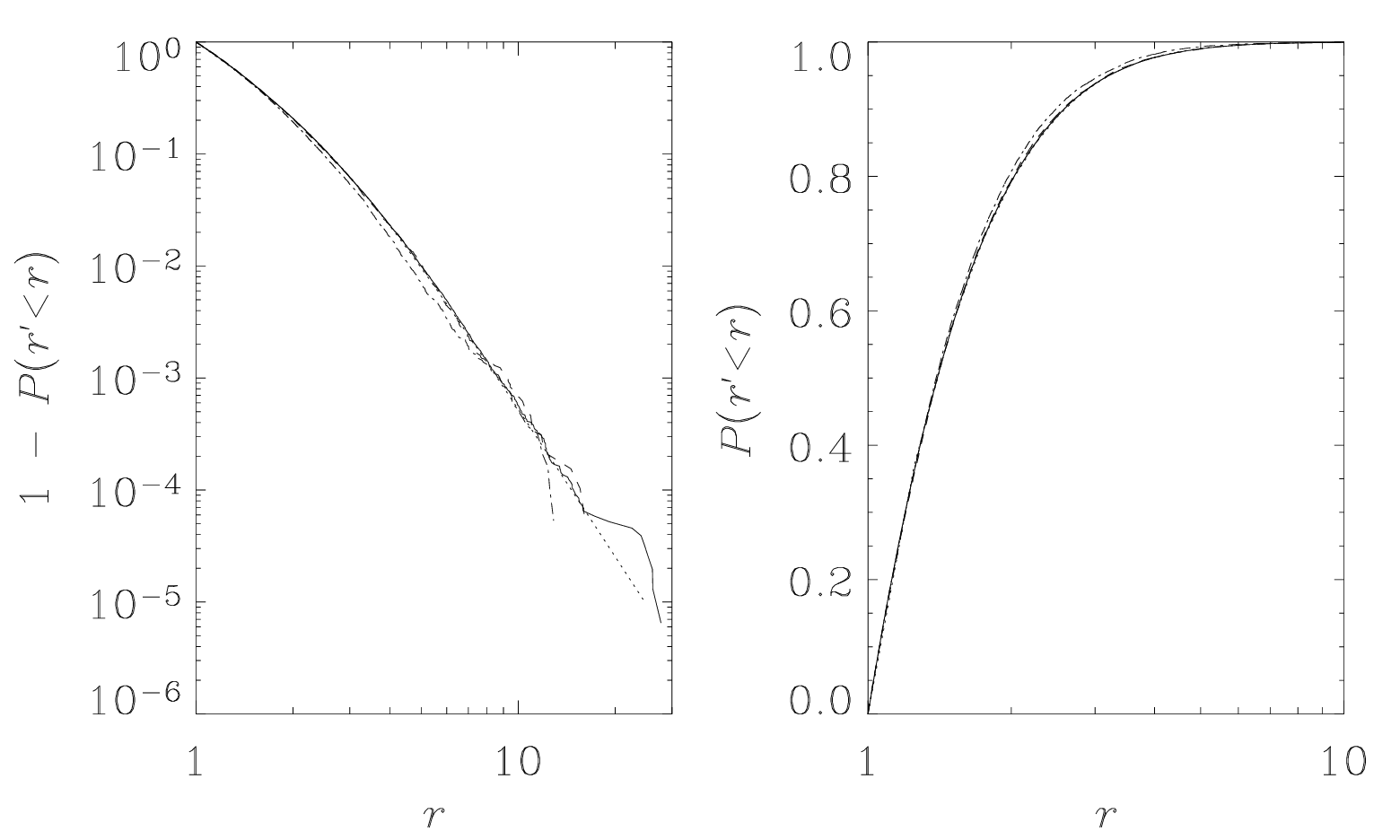}
    \end{center}
    \caption[1]{ \small The cumulative probability function $P(r^\prime<r)$
    of the ratio $r(z)$ between the peak density of a zone and its critical
    strongest link density $\rho_{sl}$, from various Poisson simulations.
    The left panel shows $1 - P(r^\prime<r)$ on a logarithmic scale, while
    the right panel shows $P(r^\prime<r)$ on a more familiar linear scale
    on $[0,1]$.  The dashed curves are drawn from haloes in a uniform
    Poisson simulation with $64^3$ particles, and the solid curves in one
    with $128^3$ particles.  The function does seem to converge as particle
    number increases.  The dotted curve (indistinguishable in the right
    panel) shows the fit in Eqn.\ (\ref{prob}).  The dot-dashed curve
    shows $P(r^\prime<r)$ of subhaloes in a large halo with density profile
    $\rho \sim r^{-2}$, Poisson sampled with $64^3$ particles.
    \label{pcdf}
    }
  \end{figure}
}

\newcommand{\grid}{
  \begin{figure}
    \begin{center}
      \leavevmode
      \epsfxsize=2in    
      \epsfbox{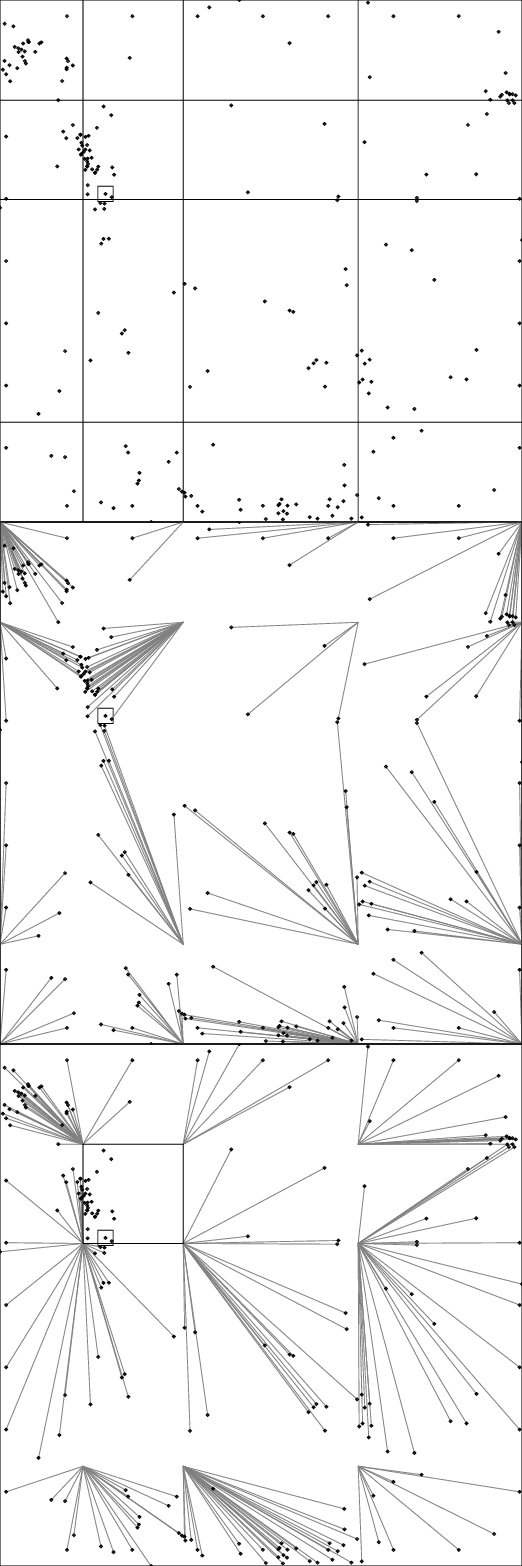}
    \end{center}
    \caption[1]{ \small Our method to find the deeper and shallower bounds
      on the potential of a halo.  The top panel shows a collection of
      particles, partitioned by a grid, spaced so that the number of
      particles in each row and column is the same.  (This is only roughly
      true, of course, if the number of particles is not divisible by the
      number of rows or columns.)  The true potential of the boxed particle
      $p$ is found by directly summing the potentials from all other
      particles.  The middle panel illustrates the method of finding the
      shallower bound, in which each particle is moved to the farthest
      corner in its cell from $p$.  The bottom panel illustrates the deeper
      bound, in which each particle is moved to the nearest corner in its
      cell to $p$, except if it is in $p$'s cell, in which case its
      potential is directly summed.
      \label{grid}
    }
  \end{figure}
}

\newcommand{\rcdf}{
  \begin{figure}
    \begin{center}
      \leavevmode
      \epsfxsize=\columnwidth    
      \epsfbox{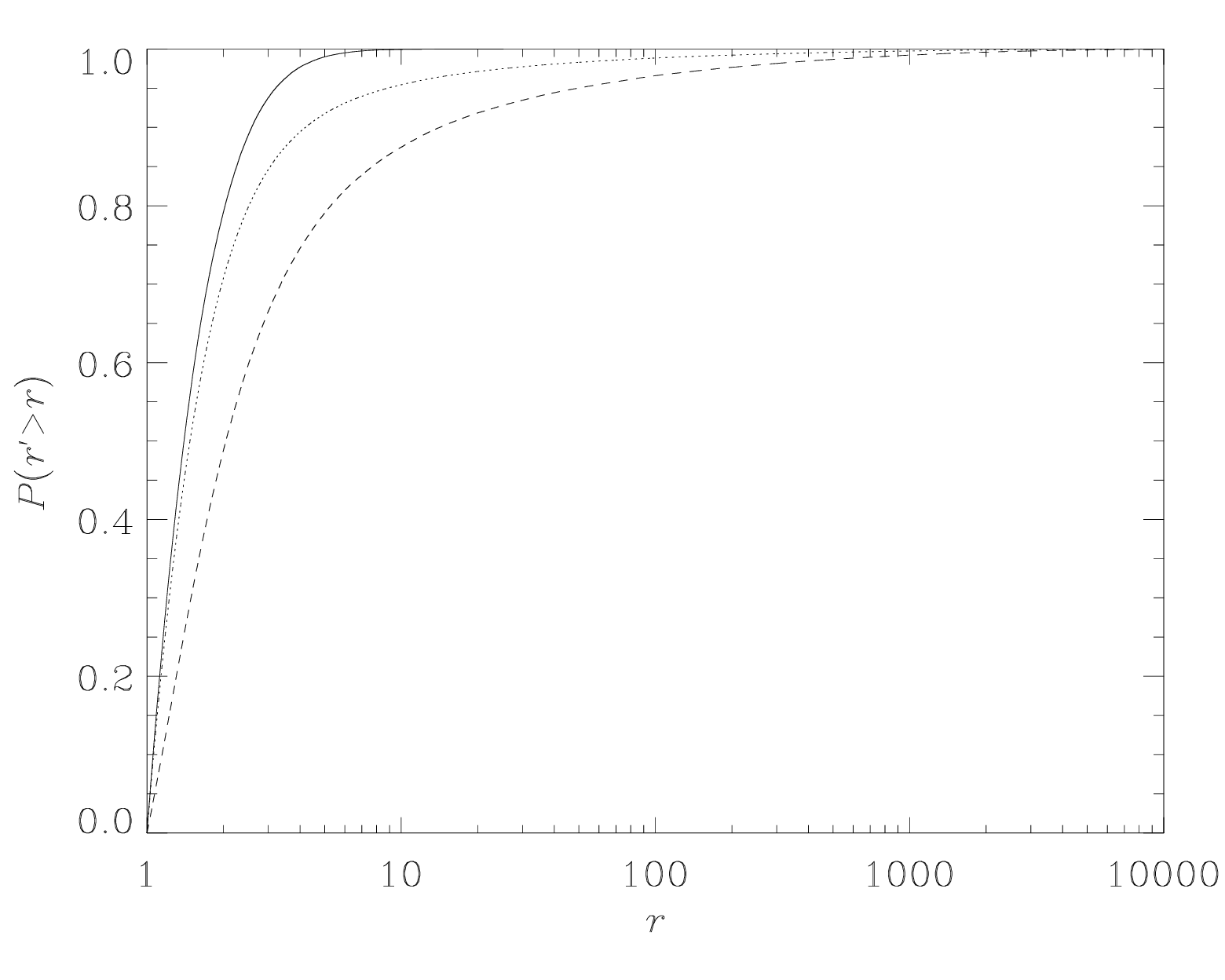}
    \end{center}
    \caption[1]{ \small The cumulative distribution function of the ratio
      $r(z)$ between the peak density of a zone and its critical strongest
      link density $\rho_{sl}$, from haloes in a uniform Poisson simulation
      (solid curve), from all haloes in the 32 $h^{-1}$ Mpc simulation
      (dotted), and from all bound haloes in this simulation (dashed).
      \label{rcdf}
    }
  \end{figure}
}

\newcommand{\crls}{
  \begin{figure}
    \begin{center}
      \leavevmode
      \epsfxsize=\columnwidth    
      \epsfbox{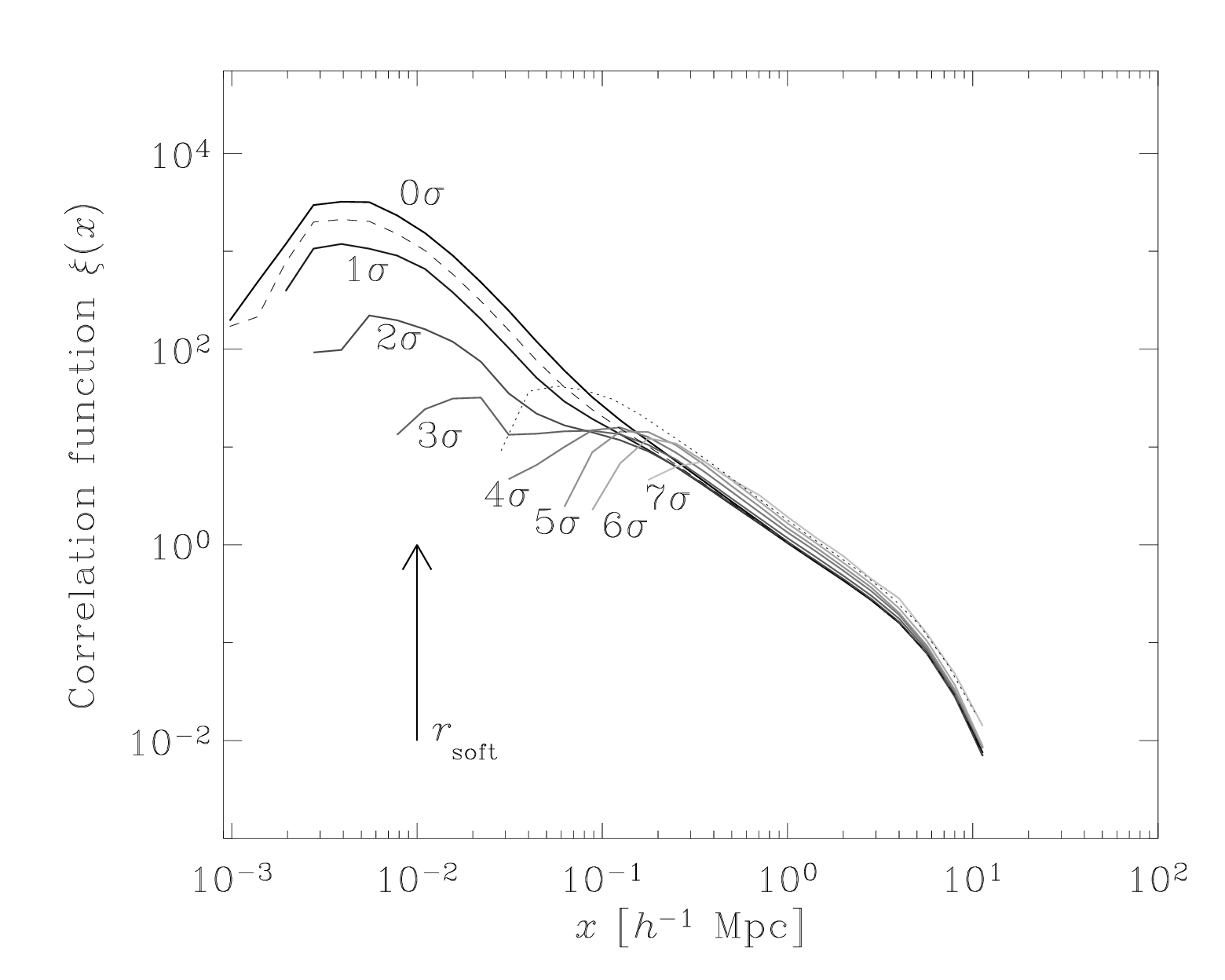}
    \end{center}
    \caption[1]{ \small Correlation functions (CFs) of haloes from a
      $256^3$ particle, 32 $h^{-1}$ Mpc $\Lambda$CDM simulation, with
      softening length $r_{soft} = 0.01$ $h^{-1}$ Mpc.  The solid lines are
      labeled with their probability cut-offs, ranging from $1-7\sigma$.
      The dashed line is a CF of all {\scshape voboz} haloes, weighting
      pairs by the product of their probabilities.  There is no cut-off
      imposed in halo size, so the smallest haloes included have only two
      particles, giving a minimum halo mass of $5 \times 10^8 M_\odot$.
      The dotted curve is the CF of all {\scshape denmax}$^2$ haloes with
      greater than 10 particles, i.e.\ with a minimum mass of $2 \times
      10^9 M_\odot$.
      \label{crls}
    }
  \end{figure}
}

\newcommand{\haloterm}{
  \begin{figure}
    \begin{center}
      \leavevmode
      \epsfxsize=\columnwidth    
      \epsfbox{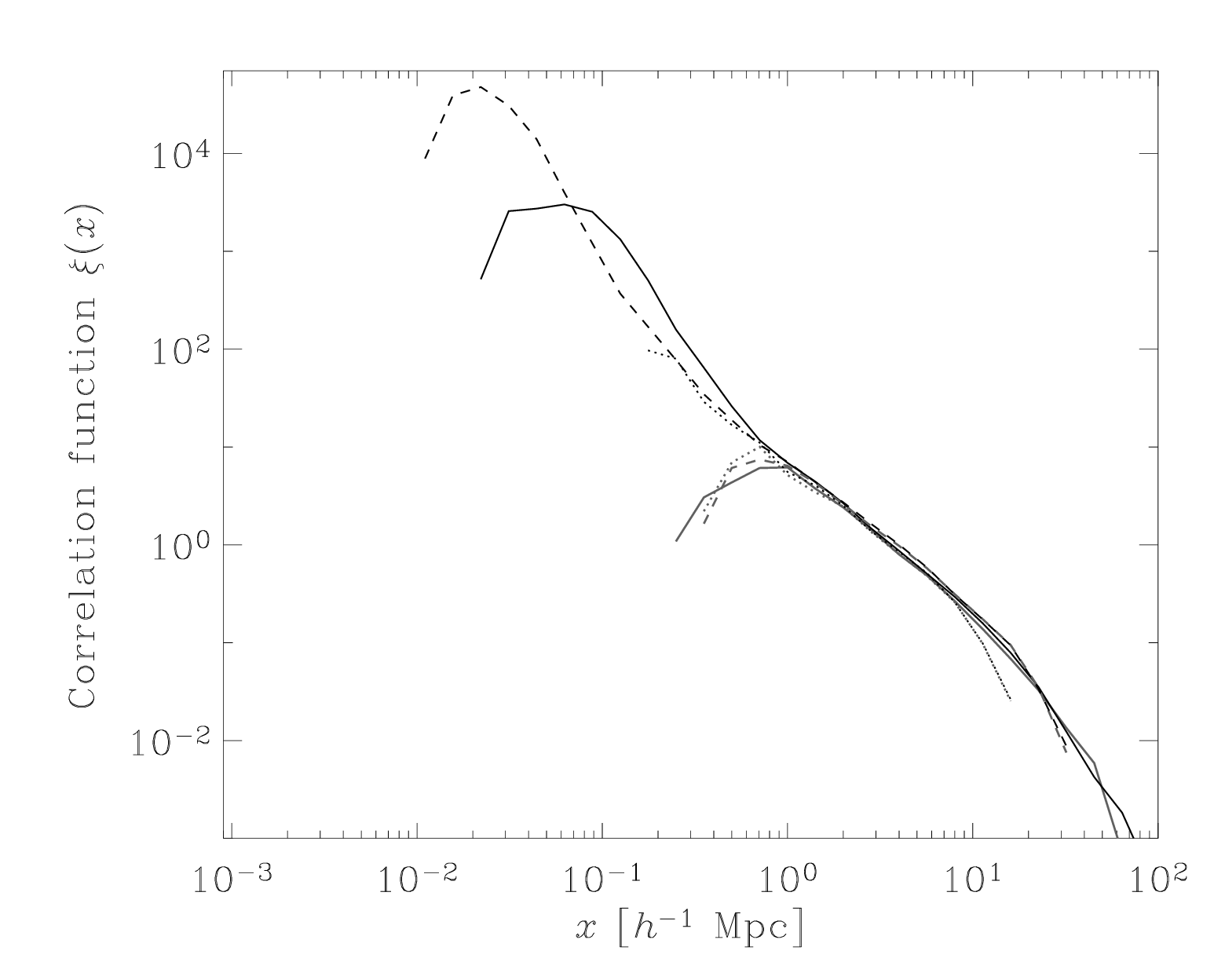}
    \end{center}
    \caption[1]{ \small An attempt to use a probability cut-off to isolate
      the `2-halo' term in the correlation function (CF) of haloes
      exceeding $10^{12}$ $h^{-1} M_\odot$ in our 256 (solid), 128
      (dashed), and 64 (dotted) $h^{-1}$ Mpc simulations.  The black curves
      are CFs of all haloes, while the grey curves are CFs of haloes
      exceeding probability cutoffs of $4\sigma$, $5\sigma$, and $6\sigma$
      in the 256, 128, and 64 $h^{-1}$ Mpc simulations, respectively.
      (C.f.\ Kravtsov et al.\ 2004, Fig.\ 8)
      \label{haloterm}
    }
  \end{figure}
}

\newcommand{\massfn}{
  \begin{figure}
    \begin{center}
      \leavevmode
      \epsfxsize=\columnwidth    
      \epsfbox{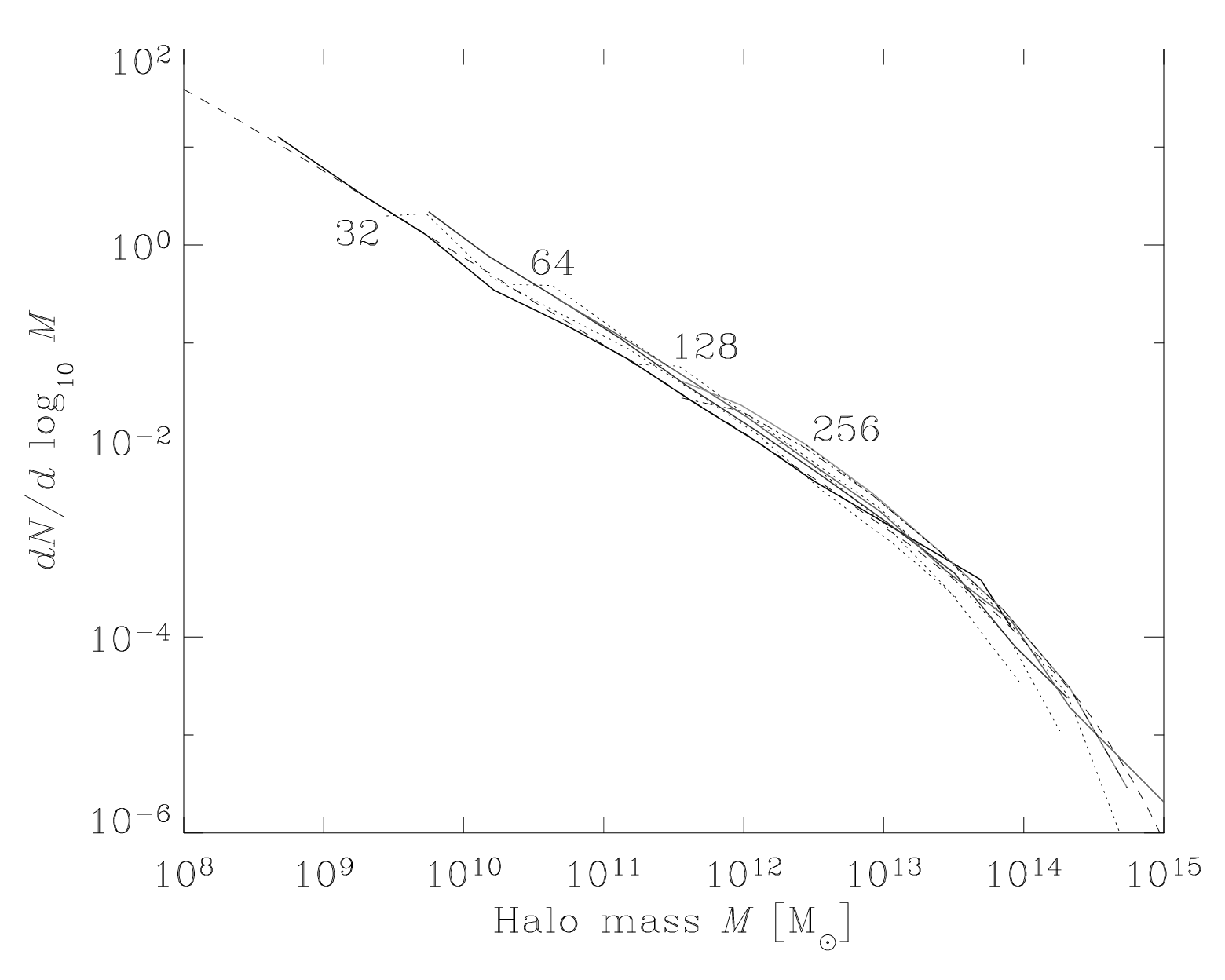}
    \end{center}
    \caption[1]{ \small Mass functions of haloes from $256^3$-particle
    $\Lambda$CDM simulations, of box size $32$, $64$, $128$, and $256$
    $h^{-1}$ Mpc.  The solid curves show {\scshape voboz} haloes, using a
    density cut-off $\rho_{min}$ of 100.  The dotted curves show haloes
    detected with {\scshape denmax}, using the canonical smoothing length
    of $1/5$ the mean interparticle separation.  The dashed line is the
    Sheth, Mo \& Tormen (2001) analytical prediction.
    \label{massfn}
    }
  \end{figure}
}

\newcommand{\massfnone}{
  \begin{figure}
    \begin{center}
      \leavevmode
      \epsfxsize=\columnwidth    
      \epsfbox{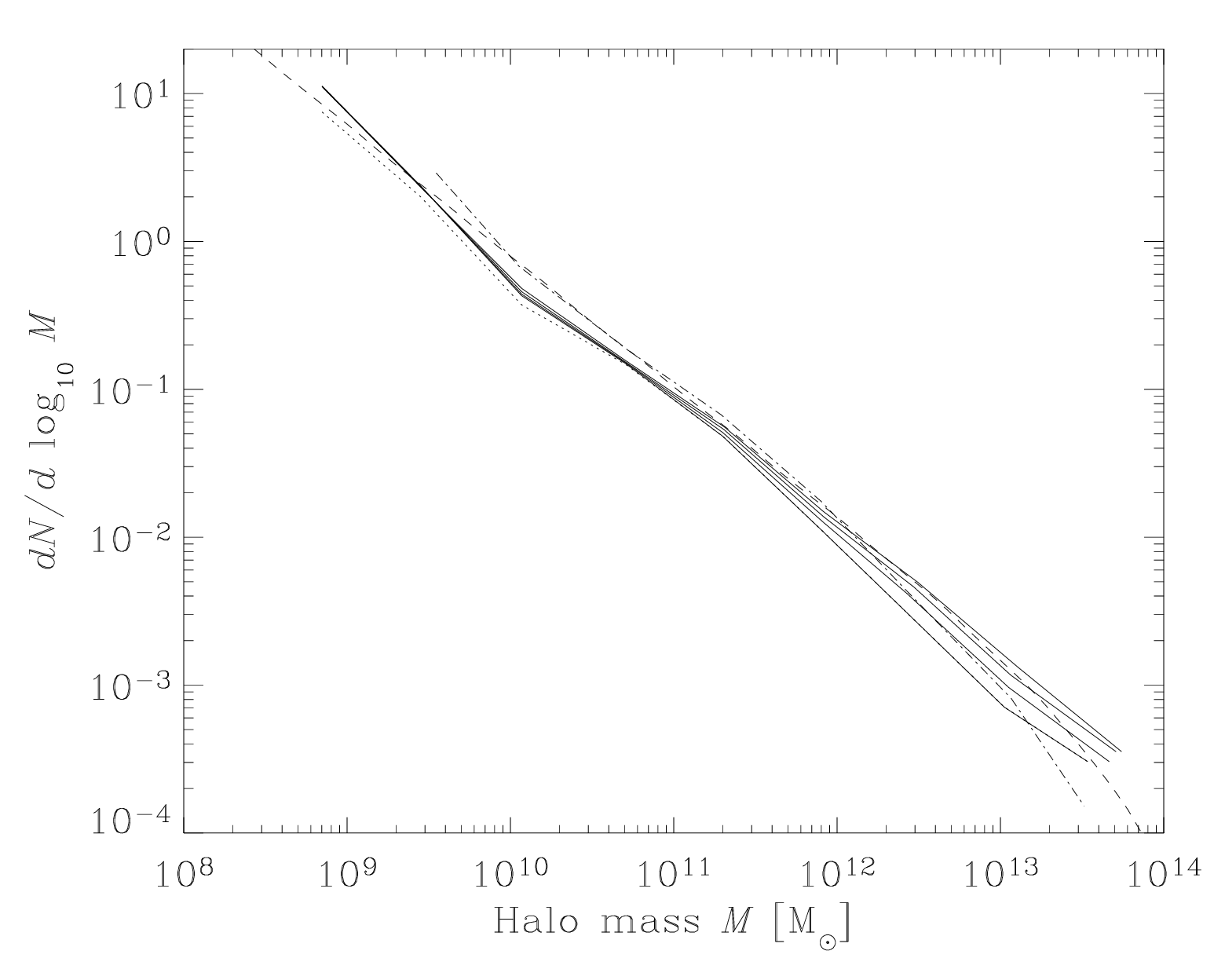}
    \end{center}
    \caption[1]{ \small Mass functions of haloes from a $256^3$-particle
    $\Lambda$CDM simulation of box size 32 $h^{-1}$ Mpc.  The solid curves
    show {\scshape voboz} haloes, using density cut-offs $\rho_{min}$ of
    (from top to bottom) 50, 100, 200, and 400.  In the dotted
    curve, we weight each halo from the $\rho_{min} = 400$ list by its
    probability; e.g.\ a 50\% probable halo is counted as half a halo.  The
    dot-dashed curve shows haloes from {\scshape denmax}, run with the
    canonical smoothing length of $1/5$ the mean interparticle separation.
    The dashed line is the Sheth, Mo \& Tormen (2001) analytical mass
    function.
    \label{massfnone}
    }
  \end{figure}
}

\newcommand{\mcomp}{
  \begin{figure}
    \begin{center}
      \leavevmode
      \epsfxsize=\columnwidth    
      \epsfbox{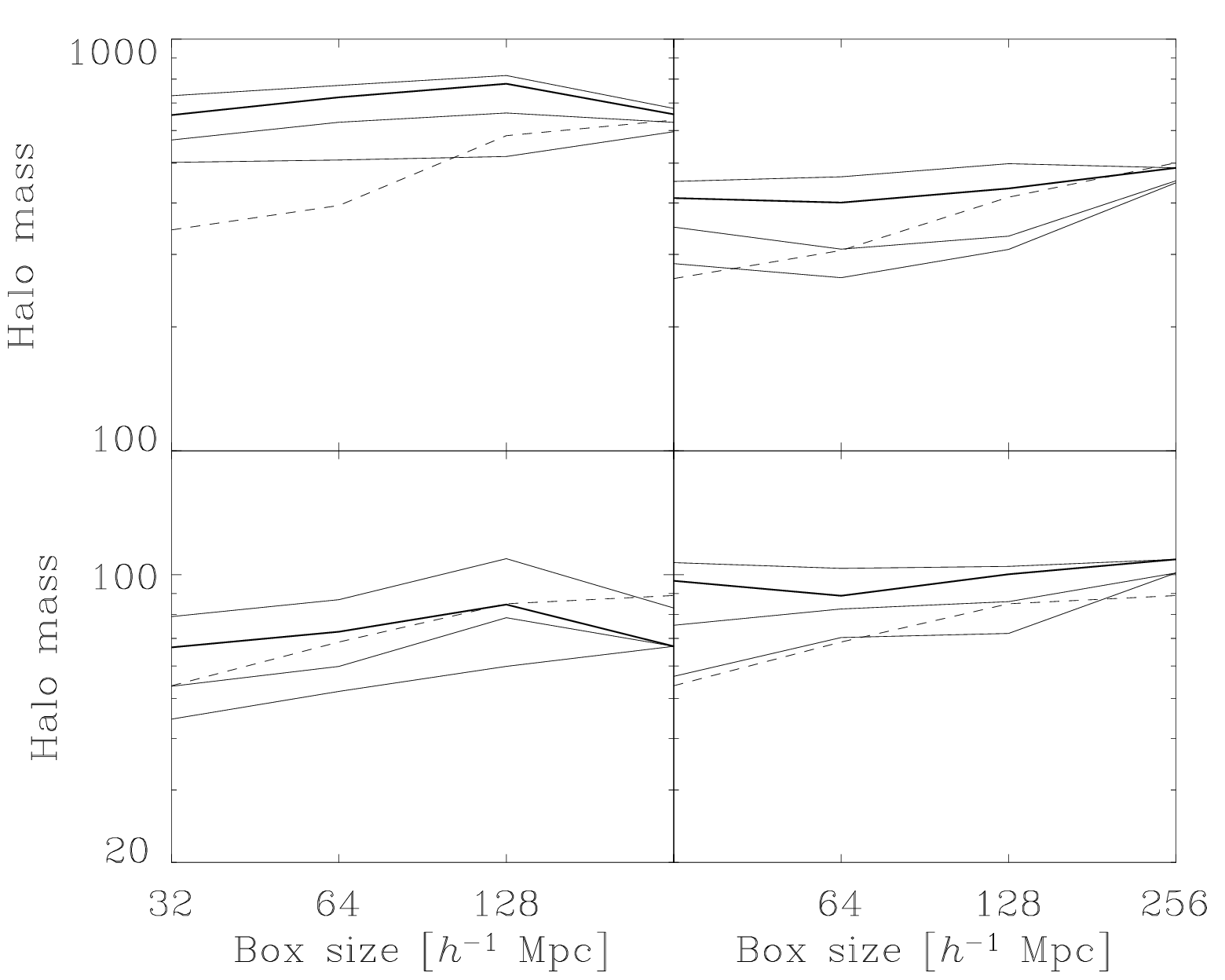}
    \end{center}
    \caption[1]{ \small Dependence of halo mass on mass resolution for four
    haloes (one in each panel) identified through all four nested
    simulations described in NHG.  The solid curves track the bound
    {\scshape voboz} halo masses (in units of particle masses in the 256
    $h^{-1}$ Mpc simulation, which is $1.2\times 10^{11} M_\odot$) in the
    four simulations using $\rho_{min} = 50$ (top), $100$ (bold), $200$,
    and $400$ (bottom).  The lines should be horizontal if the haloes are
    identical and mass resolution does not affect the halo-finding.
    Particularly for large haloes, {\scshape voboz} is much less sensitive
    to mass resolution than {\scshape denmax}.  The discreteness of the
    zones making up the smallest (bottom row) haloes becomes apparent in
    the 256 $h^{-1}$ Mpc simulation, where the zones contain only $\sim$
    100 particles.
    \label{mcomp}
    }
  \end{figure}
}

\newcommand{\ubtest}{
  \begin{figure}
    \begin{center}
      \leavevmode
      \epsfxsize=\columnwidth    
      \epsfbox{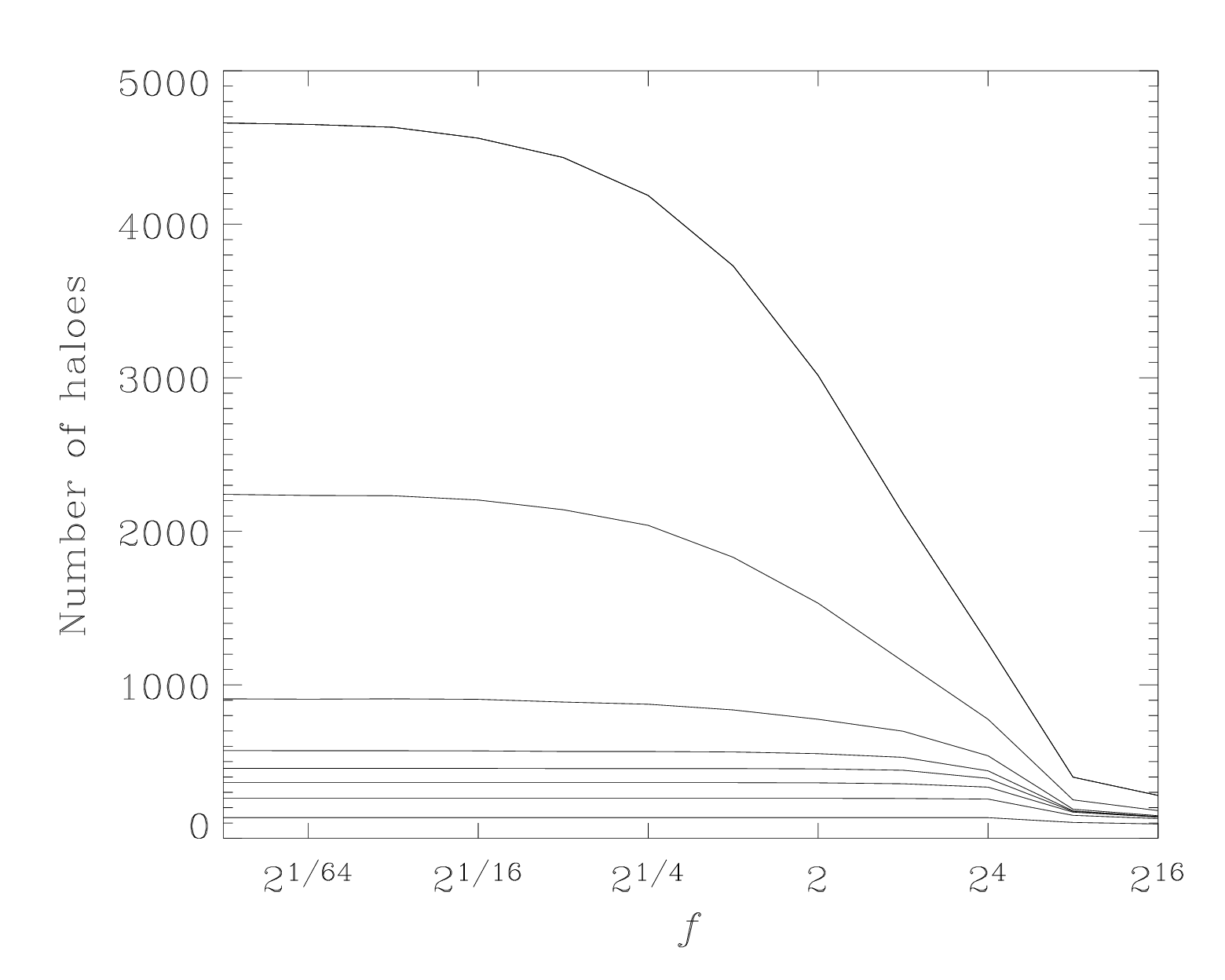}
    \end{center}
    \caption[1]{ \small Effect of unbinding coarseness on the number of
      haloes {\scshape voboz} finds in a large cluster.  The curves show
      the number of haloes satisfying various probability cut-offs, from
      $0\sigma$ (top) to $7\sigma$ (bottom).  As $f$ (the factor by which
      the potential multiplier is divided at each iteration) increases, so
      does the number of low-probability haloes that {\scshape voboz}
      completely unbinds.
      \label{ubtest}
    }
  \end{figure}
}

\newcommand{\kravmassfn}{
  \begin{figure}
    \begin{center}
      \leavevmode
      \epsfxsize=\columnwidth    
      \epsfbox{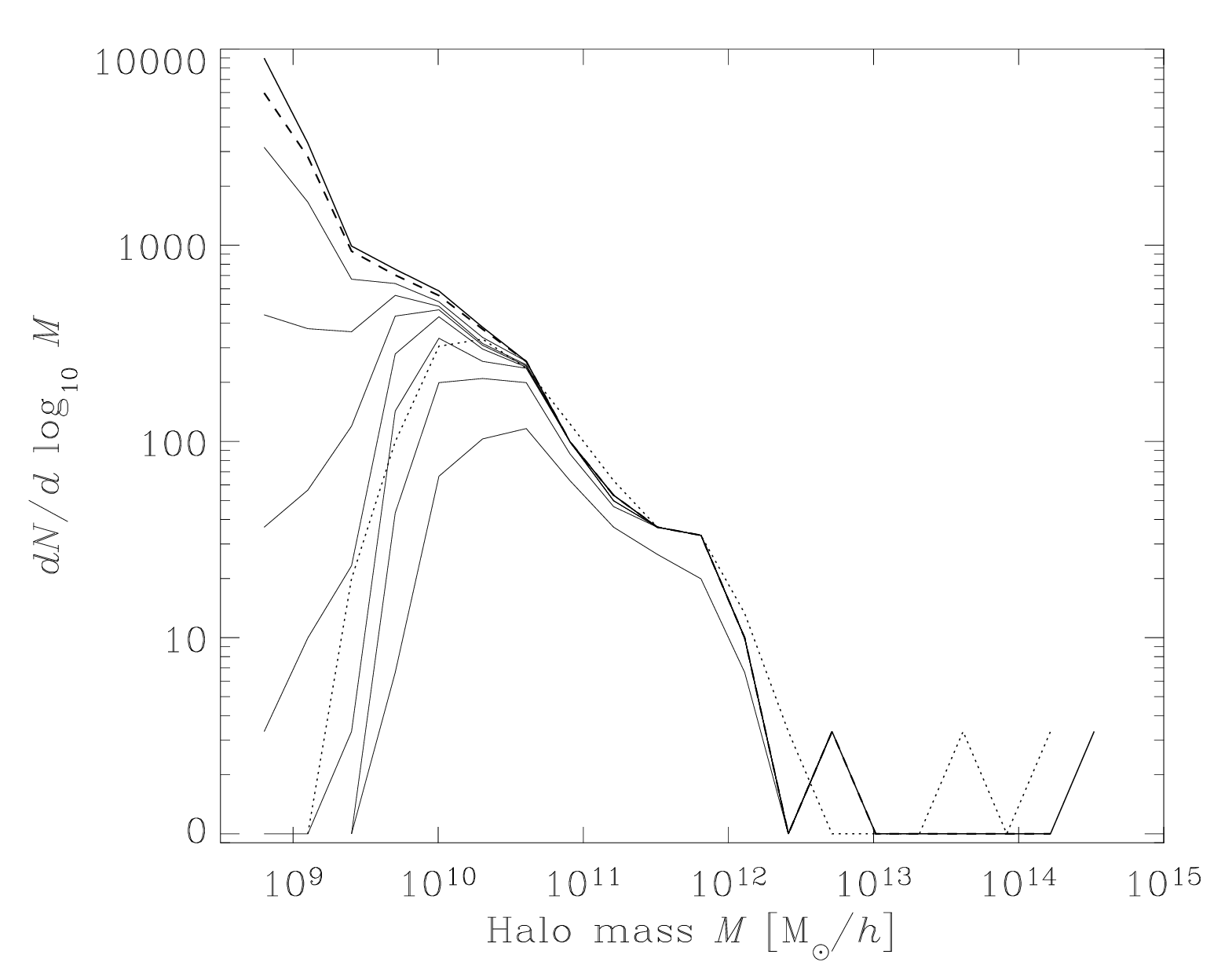}
    \end{center}
    \caption[1]{ \small Mass function of haloes from a large cluster.  The
      dotted curve shows haloes found by a variant of {\scshape bdm}.  The
      solid curves show {\scshape voboz} haloes satisfying various
      probability cut-offs.  The top (first) solid curve applies no
      cut-off, and the second applies a cut-off of $1\sigma$; the
      probability cut-off increases in increments of $\sigma$ down to the
      bottom curve, where it reaches $7\sigma$.  The dashed curve weights
      each halo with its probability.  The odd behavior at high $M$ arises
      from differing masses returned by the two HFAs for the two largest
      haloes.
      \label{kravmassfn}
    }
  \end{figure}
}

\newcommand{\kravmscat}{
  \begin{figure}
    \begin{center}
      \leavevmode
      \epsfxsize=\columnwidth    
      \epsfbox{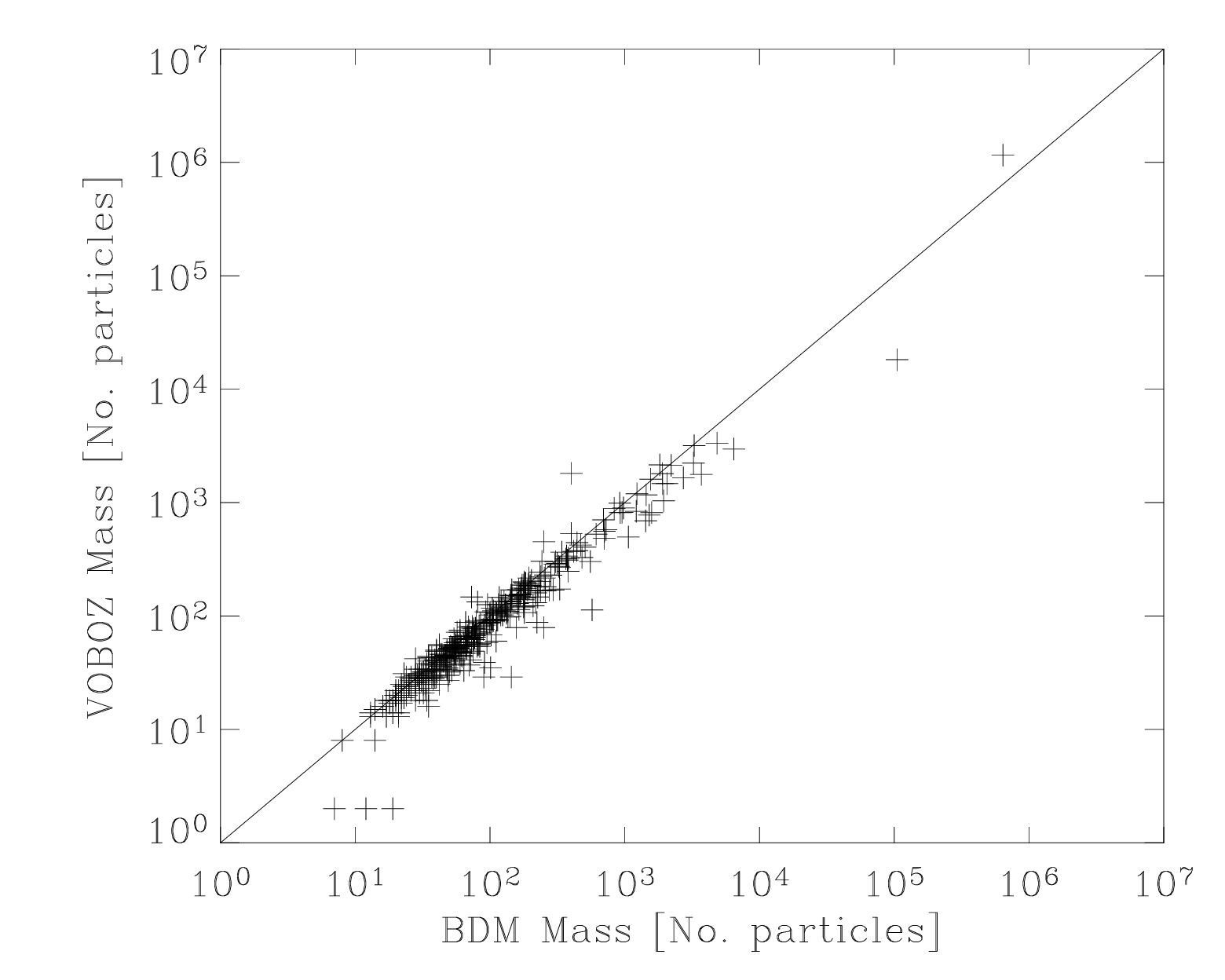}
    \end{center}
    \caption[1]{ \small Scatter plot of masses, in units of particle mass,
      of haloes detected by both a variant of {\scshape bdm}, and {\scshape
      voboz}.  The line shows where the masses should lie if the HFAs
      returned identical masses.  A {\scshape voboz} halo may be matched
      with many {\scshape bdm} haloes.  The largest halo is the cluster
      itself, which included many larger-mass particles in the actual
      simulation, so its mass is underestimated by both algorithms.
      \label{kravmscat}
    }
  \end{figure}
}
\newcommand{\weirdhalo}{
  \begin{figure}
    \begin{center}
      \leavevmode
      \epsfxsize=\columnwidth   
      \epsfbox{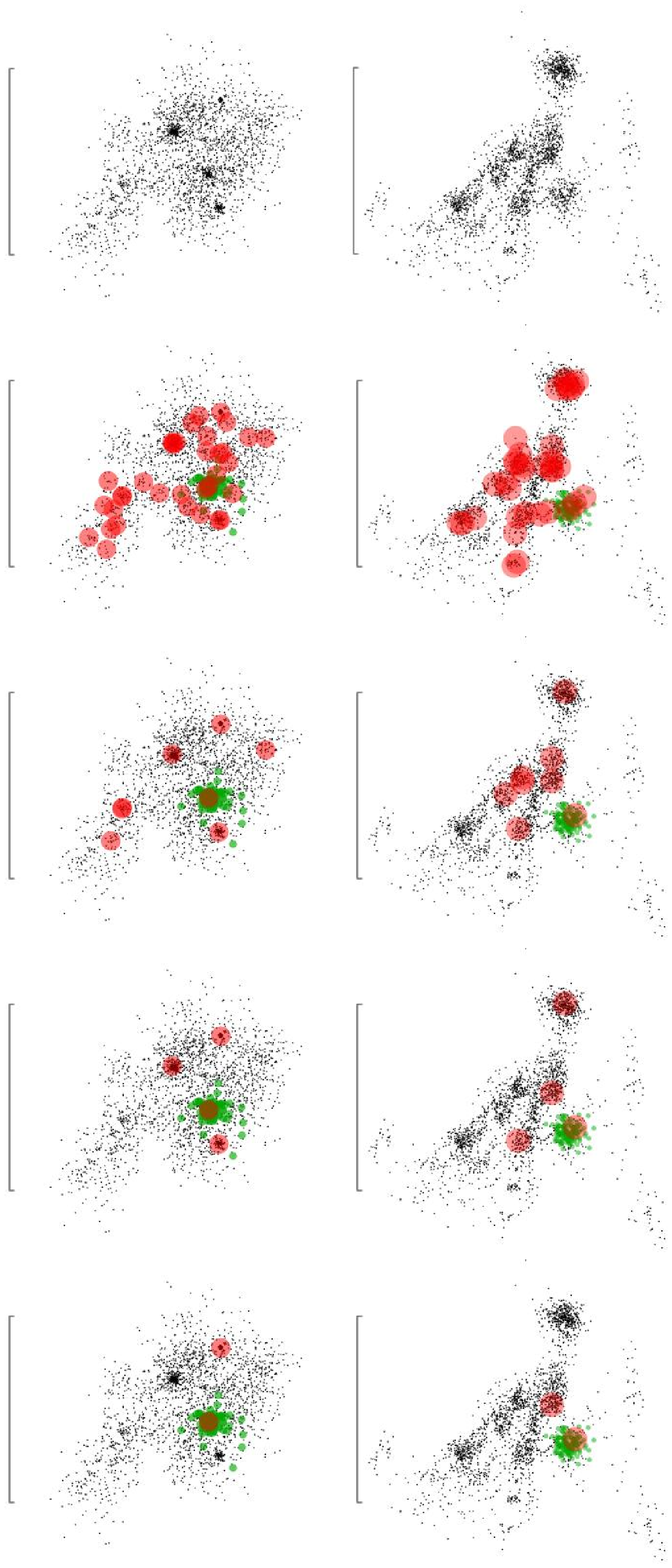}
    \end{center}
    \caption[1]{ \small A particularly messy `halo' in a large cluster.  On
      the left, the `halo' is shown in position space (with a 0.8 $h^{-1}$
      Mpc bracket); on the right, it is shown in velocity space (with a
      1000 km/s bracket).  The black dots represent all particles in the
      `halo,' including those eventually unbound; the raw particles appear
      in the first row.  In subsequent rows, those particles which
      {\scshape voboz} deems bound to the final halo are replaced with
      bloated green particles.  In position space, the large red circles
      show the centers of haloes which lie within this larger `halo;' in
      velocity space, the circles show their velocity centroids.  The
      second row shows all haloes; the third row shows all haloes with
      probabilities above $2\sigma$; the fourth shows haloes above
      $4\sigma$; and the bottom shows haloes above $7\sigma$.  This figure
      was produced using Nick Gnedin's {\scshape ifrit} visualization tool,
      at \url{http://casa.colorado.edu/~gnedin/IFRIT/}.
      \label{weirdhalo}
    }
  \end{figure}
}

\begin{document}
\title[VOBOZ: An Almost-Parameter-Free Halo-Finding Algorithm]
      {VOBOZ: An Almost-Parameter-Free Halo-Finding Algorithm}

\author[Mark C.\ Neyrinck, Nickolay Y.\ Gnedin, and Andrew J.S.\ Hamilton]
{Mark C.\ Neyrinck$^{1,3}$, Nickolay Y.\ Gnedin$^{2,3}$, and Andrew J.S.\ Hamilton$^{1,3}$ \\
  $^{1}$JILA, University of Colorado, Boulder, CO 80309\\
  $^{2}$Center for Astrophysics and Space Astronomy, University of Colorado, Boulder, CO 80309\\
  $^{3}$Department of Astrophysical and Planetary Sciences, University of Colorado, Boulder, CO 80309 \\
  email: {\tt Mark.Neyrinck@colorado.edu}}
\date{2004 July 6}
\pubyear{2004}

\maketitle
  
\begin{abstract}
  We have developed an algorithm to find haloes in an $N$-body dark matter
  simulation, called {\scshape voboz} ({\scshape vo}ronoi {\scshape bo}und
  {\scshape z}ones), which has as little dependence on free parameters as
  we can manage.  By using the Voronoi diagram, we achieve nonparametric,
  `natural' measurements of each particle's density and set of neighbors.
  We then eliminate much of the ambiguity in merging sets of particles
  together by identifying every possible density peak, and measuring the
  probability that each does not arise from Poisson noise.  The main halo
  in a cluster tends to have a high probability, while its subhaloes tend
  to have lower probabilities.  The first parameter in {\scshape voboz}
  controls the subtlety of particle unbinding, and may be eliminated if one
  is cavalier with processor time; even if one is not, the results saturate
  to the parameter-free answer when the parameter is sufficiently small.
  The only parameter which remains, an outer density cut-off, does not
  influence whether or not haloes are identified, nor does it have any
  effect on subhaloes; it only affects the masses returned for supercluster
  haloes.

\end{abstract}

\begin {keywords}
  methods: $N$-body simulations -- large-scale structure of Universe --
  galaxies: haloes -- cosmology: theory -- methods: data analysis --
  galaxies: formation.
\end {keywords}

\section{Introduction}

A crucial step in comparing $N$-body simulations to the observed galaxy
distribution is to identify the possible sites of galaxy formation, called
dark-matter haloes, in the simulations.  Unfortunately, the concept of a
dark-matter halo is not precisely defined.  There is no firm observational
definition of dark-matter haloes, since they can only be observed
indirectly; for example, through gravitational lensing.  There are a couple
of possible theoretical definitions.  One of them is a region exceeding a
certain overdensity, such as the canonical overdensity of virialization,
200.  This is often used when seeking Halo Occupation Distributions (e.g.\
Berlind \& Weinberg 2002), which statistically characterize the number of
galaxies inside haloes (implicitly dark-matter hereafter) as a function of
halo mass.  However, if we want to look beyond the statistical placement of
galaxies inside haloes, we should use another definition of a halo (or
subhalo): a density peak to which some mass is gravitationally bound.  In
the language of an $N$-body simulation, a particle is the core of a halo if
it is a local density maximum, and there exists at least one other particle
bound to it.

One of the first halo-finding algorithms (HFAs), still in wide use because
it is so fast and conceptually simple, is the Friends-of-Friends algorithm
(Davis et al.\ 1985).  This HFA groups together all particles within a
specified linking length, a free parameter which is usually set by the
canonical overdensity of virialization.  Friends-of-Friends is useful if
one is looking for large structures exceeding this overdensity, but it is
incapable of finding subhaloes within these structures, and sometimes
structures are unduly linked if there happens to be a stream of particles
connecting them.

Most HFAs developed since Friends-of-Friends begin with an explicit
measurement of the density, which is not uniquely or obviously defined
given a set of particles.  In one Eulerian method ({\scshape denmax},
Bertschinger \& Gelb 1991), each particle is smoothed with a Gaussian of a
fixed spatial resolution.  As with the Friends-of-Friends algorithm, the
free parameter is set roughly by the critical overdensity of virialization.
While this value of the parameter tends to give virialized objects, it
smears out subhaloes (Neyrinck, Hamilton \& Gnedin 2004, hereafter NHG);
one runs the risk of missing structures smaller than any fixed smoothing
length.  On the other hand, using a smoothing length that is too small
misses the less-dense outskirts of haloes.  Another HFA, called {\scshape
bdm} (Klypin \& Holtzman 1997), finds density maxima by placing spheres
randomly in the simulation, and then moving them at each iteration to the
center of mass of particles within them.  Maxima are then joined if they
lie within a specified radius.  Another way to find the density, called
{\scshape skid} (Weinberg, Hernquist \& Katz 1997, Jang-Condell \&
Hernquist 2001) uses a Lagrangian, `smoothed particle hydrodynamics' (SPH)
density estimate based on the distances to the nearest $N_{dens}$
particles.  This density estimate is arguably an improvement over {\scshape
denmax}'s because there is no fixed spatial resolution, but in its place
there is an arbitrary, fixed mass resolution.  This is undesirable because
haloes can exist with only two particles, which a fixed mass resolution is
likely to miss.

The next step in halo finding is to group the particles together.  In
{\scshape denmax} and {\scshape skid}, particles slide along density
gradients until they reach density maxima.  In {\scshape hop} (Eisenstein
\& Hut 1998), which uses a Lagrangian density estimator similar to that in
{\scshape skid}, each particle `hops' to the densest particle among its
neighbors, and continues in this manner until it reaches a local density
maximum.  Then, groups of particles are joined together if saddles linking
them exceed a specified density, another free parameter.  Recently, Kim \&
Park (2004) have developed another HFA, called {\scshape psb}.  Around each
density maximum (calculated with a small spatial smoothing length),
{\scshape psb} finds the largest isodensity contour enclosing only that
peak, using a couple of parameters to do so.  It then assigns neighboring
particles to density peaks using considerations such as whether they are
energetically bound, and whether they lie outside the tidal radius.  It
seems that {\scshape psb} reliably uncovers subhaloes, but it does require
several free parameters.

\section{Method}

Our HFA, {\scshape voboz} ({\scshape vo}ronoi {\scshape bo}und {\scshape
z}ones), identifies haloes in three steps: (1) measuring the density at
each particle, (2) grouping sets of particles around density maxima which
plausibly form a halo using only spatial information, and (3) unbinding
particles from haloes if their velocities exceed the escape velocity from
their halo at their position.

{\scshape voboz} also performs a measurement new to the world of HFA's: it
measures the probability that each halo did not arise from Poisson noise.
Such a measurement is quite useful for subhaloes within haloes which may or
may not exist.  Previous HFA's either identify or do not identify
questionable haloes; the results do not indicate haloes which were just
barely identified, nor haloes which barely escaped detection.

\subsection{Calculation of the Voronoi Diagram}
The method we use to calculate the densities of particles employs the
Voronoi diagram (VD), a unique, nonparametric tessellation of a space
containing particles.  Ideas related to the VD have existed for centuries,
but Voronoi (1908) introduced it in its modern form.  The standard
reference is Okabe et al.\ (2000), in which appears a survey of Voronoi
applications in an extensive array of fields including biology, forestry,
archaeology, urban planning, and meteorology.  A reference with a view
toward astronomy is provided by van de Wegaert (1994), who, with his
collaborators, has championed the use of Voronoi methods in various
contexts related to large-scale structure and cosmology, starting with the
description of voids as polyhedra (Icke \& van de Weygaert 1987).  A major
current astronomical application of Voronoi methods is in identifying
clusters of galaxies from surveys (e.g.\ Ebeling \& Wiedenmann 1993,
Ramella et al.\ 2001, Kim et al.\ 2002, Marinoni et al.\ 2002).  The
Delaunay Tesselation Field Estimator (Schaap \& van de Weygaert 2000),
essentially the same as our density estimator (it uses the dual of the VD,
the Delaunay tesselation), has been shown to be superior to SPH (Pelupessy,
Schaap \& van de Weygaert 2003, Schaap 2004).  Quite recently, Arad, Dekel
\& Klypin (2004) used the DTFE to explore the properties of 6-D
position-velocity space in an $N$-body simulation.

The VD of a set of particles $P$ is defined as follows.  The Voronoi {\it
cell} $V(p_i)$ around a particle $p_i$ in $P$ is the interior of the
polyhedron of points closer to $p_i$ than to any other particle.  The most
intuitive way to find a Voronoi cell around $p_i$ is, for each other
particle $p_j$ in $P$, to put up planes perpendicularly bisecting the line
segments connecting $p_i$ and $p_j$.  $V(p_i)$ will then be the polyhedron
formed by these bisecting planes which contains $p_i$.  Figure \ref{lines}
shows a two-dimensional Voronoi diagram of a set of particles from an
$N$-body simulation.  A Voronoi {\it neighbor}, or {\it adjacency}, of a
particle $p_i$ is a particle $p_j$ such that $V(p_j)$ borders $V(p_i)$; the
set of Voronoi neighbors of $p_i$ demarcate $V(p_i)$.

\lines

We may then define the density of a particle as $1/\rm{Volume}$$(V(p_i))$.
Not only does the VD give a unique density, with no free parameters, and
with infinite spatial resolution, but it gives, arguably, the most local
density estimate that contains meaningful information, uncovering every
possible morsel of structure.  It also returns a `natural' set of neighbors
for each particle, with no fixed mass resolution.  In addition, it benefits
from having been studied extensively.  Many statistical properties of
Poisson Voronoi diagrams (VD's applied to Poisson point processes) are
well-known, and sometimes are even analytically calculable.  For example,
the average number of neighbors of a particle in a three-dimensional
Poisson VD is $(48\pi^2/35) + 2 \approx 15.535$, with a standard deviation
of 3.32 (Okabe et al.\ 2000).

To calculate the VD, we used the Quickhull algorithm (Barber, Dobkin \&
Huhdanpaa 1996), which runs in $O(n \log n)$ time, where $n$ is the number
of input points.  We also used the Qhull (the implementation of Quickhull)
package to compute volumes of polyhedra.  Because Qhull's memory
requirements become prohibitive when it is run directly on a large set of
particles ($\sim$ 10 GB for a 128$^3$-particle box), and also to produce
intermediate outputs (valuable in interrupted runs), we split the box on
which the VD is calculated into sub-boxes.  A user constrained by memory
may have to split the box into many pieces for them to fit into memory.

It is important to check that dividing the box does not alter particles'
volumes and sets of neighbors.  See the appendix for details of how we do
this.

\subsection{Zones}\label{zones}
We believe that the VD gives the best way of finding densities and
adjacencies in a set of particles, and a prospective user of {\scshape
voboz} may wish to use it for this alone.  Our method to group particles
was more empirically found, but it does seem to work quite well in
detecting even the smallest structures, and also in efficiently mapping out
larger structures.

It is easy to define density maxima among particles given their densities
and neighbors: a particle is a density maximum if its density is higher
than any of its neighbors'.  To find particles which may belong to that
particle's halo, we send each particle in the set to its neighbor with the
highest density, and repeat this process until every particle is at a
density maximum, a procedure similar to that in {\scshape hop}.  We define
a density maximum's {\it zone} to be the set of particles which jump to it;
we define a zone's {\it peak} to be its density maximum.  Figure
\ref{zoning} shows this process.

\zoning

However, we are far from finished with our analysis.  In a Poisson VD, we
found that $1/13.6$ of the particles were density maxima.  In our
large-scale structure simulations, we typically found that about $1/19$ of
the particles were density maxima.  This is not an overwhelming difference,
so many of the density maxima seem to come from Poisson noise.  `Zoning'
provides a convenient, fast partition of the set of particles, but zones
can be oddly shaped, not necessarily resembling idealized mountains around
peaks.  This is for a few reasons: particles' spatial positions are not
explicitly considered in zoning, only their densities and adjacencies.
Also, the fact that the peak of a particle's zone lies on the steepest path
up the density slope does not mean that there is no other zone to which the
particle could conceivably belong.  For example, if a particle is a local
density minimum, it could be argued that its first jump could be to any of
its neighbors.  Many zones turn out to be fake, and the ones which lie at
the centers of real haloes may contain far fewer particles than they
should, because the haloes have been partitioned into many, often spurious,
zones.  However, we contend that it is a good thing to partition the
particles into potential haloes in the finest (i.e.\ least coarse) possible
way, which this arguably accomplishes.

It is therefore necessary to join some zones together; we do so referring
to the following intuitive aid, illustrated in Fig.\ \ref{tank}.  Imagine
that the particles occupy a two-dimensional surface with height given by
their densities, enclosed in a water tank.  We perform the following
procedure for each zone $z$: Fill the tank until the peak of $z$ is
submerged, and then drain the water until a path emerges from the peak of
$z$ to the peak of another zone.  The lowest-density particle on that path,
with density $\rho_{sl}$, is the `strongest link' $z$ has with any
neighboring zone.  The strongest link is the highest-density particle in
the set of lowest-density particles on each path from the peak of $z$ to
the peaks of adjacent zones.  In practice, this particle is quickly found,
since we only need to search the border particles of each zone.

\tank

We continue adding adjacent zones to $z$ until a zone joins whose peak
exceeds $z$'s in density; we call the end product a halo.  The zones are
not always joined one-at-a-time; to continue the analogy with the water
tank, the strongest link density $\rho_{sl}$ decreases monotonically, often
uncovering multiple zones at once.  Thus, all zones connected to $z$ by
links with densities exceeding $\rho_{sl}$ are considered as a single unit
to merge with $z$, and the existence of one particle in this collection of
zones with density higher than that of $z$'s peak will break the merger.

We have not yet discussed a criterion to halt the growth of large haloes
well out into voids; without one, the zone with the densest particle in the
simulation would grow to encompass the entire simulation.  We tried to find
some mathematical criterion to stop unambiguously at the edge of a halo,
but had little success; everything we tried was easily tricked by complex
geometries.  One reason for the unbinding process, discussed in \S
\ref{unbinding}, is to mitigate this effect.  Although there could exist a
foolproof, parameter-free halo edge-detection method which has eluded us
(and which, with any luck, will find its way into {\scshape voboz} version
2.0), we reluctantly introduce a parameter $\rho_{min}$ to limit the
descent of $\rho_{sl}$, so that haloes do not grow across voids.  The
presence of a parameter is inherently undesirable, but it only affects the
size of the largest haloes which are unquestionably present, and not the
masses and identification of subhaloes, which is where the most headway is
to be made in HFA's.  We will discuss this parameter further in \S
\ref{masstests}.  We should also note that with haloes defined in this way,
particles may belong to multiple haloes, e.g.\ to both a subhalo and its
parent halo.  While this is a matter of convention, we believe it is better
to include subhaloes in the masses of their parents to which they
gravitationally bound, and not to excise them simply because they are bound
by themselves as well.

\pcdf

In other HFAs with Lagrangian density estimators like {\scshape hop}, many
spurious zones arise as in {\scshape voboz}, but the problem is not
directly addressed; zones are merely merged together if they exceed a
density cut-off, and the substructure is forgotten.  We use a different
philosophy in {\scshape voboz}: we measure for each halo or subhalo the
probability that it exists, i.e.\ that it did not arise from Poisson noise.

This probability is judged according to the ratio $r(z)$ of the peak
density of $z$ to the critical $\rho_{sl}$ of the particle linked to a zone
with a higher-density peak, halting $z$'s growth.  We can turn this ratio
into a probability by subjecting a Poisson set of particles to the same
algorithm, and forming a cumulative distribution of its $r(z)$'s.  The
probability that a halo is real is then $P(r) = P_{\rm Poisson}(r^\prime
{<} r)$, where $P_{\rm Poisson}$ is drawn from the Poisson realization.
Figure \ref{pcdf} shows $P(r)dr$, the probability that a halo is fake as a
function of its ratio $r(z)$.  The dotted line in Fig.\ \ref{pcdf} is a fit
we have made to $P_{\rm Poisson}(r^\prime < r)$:
\begin{equation}  
  1 - P_{\rm Poisson}(r^\prime < r) = \frac{1.077}{r^{1.82}+0.077r^{4.41}}
  \label{prob}
\end{equation}
Table \ref{ratios} shows the ratios, calculated using Eqn.\ (\ref{prob}),
which correspond to values of the standard Gaussian sigma: $1\sigma$
corresponds to a probability of 68.3\%, $2\sigma$ to 95.4\%, $3\sigma$ to
99.7\%, etc.  By setting equal the two terms in the denominator of the fit
expression, we find that there is a break where the power law steepens at
$r \approx 2.69$ ($1.7\sigma$).  This might then be a good candidate for a
natural cut-off in $r$ to separate real from fake haloes, with no reference
to the type of data being analyzed.  However, we do not recommend imposing
a strict, {\it a priori} cut-off in $r$.

\begin{table*}
  \caption{ \small Ratios $r(z)$ between a zone's peak and its strongest
    link density, corresponding to probabilities $P(r^\prime<r)$,
    calculated using Eqn.\ (\ref{prob}).}
  \label{ratios}
  \begin{tabular}{lll}
  No.\ $\sigma$ & $1-P(r^\prime<r)$ & $r(z)$\\
    0 & 1 & 1\\
    1 & 0.317 & 1.69\\
    2 & $4.55 \times 10^{-2}$ & 3.30\\
    3 & $2.70 \times 10^{-3}$ & 6.82\\
    4 & $6.33 \times 10^{-5}$ & 16.3\\
    5 & $5.73 \times 10^{-7}$ & 47.3\\
    6 & $1.97 \times 10^{-9}$ & 171\\
    7 & $2.56 \times 10^{-12}$ & 773\\
\end{tabular}
\end{table*}

It might behoove us to test this fit with an even larger Poisson
simulation, but it is academic to test whether a halo is fake with a
$10^{-4}$ or $10^{-5}$ probability; the probabilities matter the most when
a halo is questionable.  It is perhaps more relevant to see what happens if
a non-uniform distribution (such as that in an $N$-body simulation) is
Poisson-sampled.  We put a single giant halo with a density profile $\rho
\sim r^{-2}$ in a periodic $64^3$-particle box by giving each particle a
uniformly random distance to the center, and a uniformly random pair
$(\cos\theta,\phi)$ of angles.  The cumulative $r(z)$ distribution of the
fake haloes detected in this simulation appears as the dot-dashed line in
Fig.\ \ref{pcdf}; the distribution is reassuringly similar to that from the
uniform Poisson distribution.

\subsection{Unbinding}\label{unbinding}

We now have a set of prospective haloes defined using particles' spatial
positions.  However, these prospective haloes may include particles which
are not physically bound; for example, a portion of a nearby filament may
be mistakenly included.  We therefore test particles for boundness to their
halo(es), for the first time introducing velocity information.  For each
halo, we compare the kinetic and potential energies of each particle,
equivalent to comparing a particle's velocity to its escape velocity.  This
does not always correctly predict whether the particle will be bound to the
halo in the future, but it is a good estimate.  The unbinding process is
iterative, i.e.\ unbound particles are not included in the next iteration's
unbinding calculations for other particles.

Dividing out the mass of the particle, the kinetic energy of particle $i$
is $\frac{1}{2} |(\mathbf{v}_i-\mathbf{v}_c) + H(\mathbf{x}_i -
\mathbf{x}_c)|^2$, where $\mathbf{v}_c$ is the velocity centroid of the
original zone of the halo, $H$ is the Hubble constant, and $\mathbf{x}_c$
is the position of the central particle of the halo.  We use the velocity
centroid of only the original zone because {\scshape voboz} sometimes joins
together large regions which are unrelated in velocity space, which could
skew the velocity centroid if the entire halo were included in the average.
This is discussed further at the end of \S \ref{krav}.

\grid

We calculate the potential at each particle $i$ directly:
\begin{equation}
  \Phi(\mathbf{x}_i) = G\sum_{j \not= i}\frac{M_j}{|\mathbf{x}_j -
  \mathbf{x}_i|},
\label{potential}
\end{equation}
where $j$ is summed over all bound particles in the halo, and $M_j$ is the
mass of the $j$th particle.  However, for a halo with many particles, this
direct, $O(n^2)$ calculation of the potential for all particles is
unwieldy.  For this reason, we find shallower and deeper bounds for the
potential which are calculable in less time.  If a particle is bound in the
shallower potential, it is bound in the real potential; if it is unbound in
the deeper potential, it is unbound in the real one.  The true potential is
only calculated if the particle is unbound in the shallower potential but
bound in the deeper one.

To get these bounds, we partition the halo on a three-dimensional grid with
unequal spacing (see Fig.\ \ref{grid}).  In each dimension, the grid
spacing is set so that the number of particles in each one-dimensional bin
is the same.  In two dimensions, as in Fig.\ \ref{grid}, this would mean
that the number of particles in each row and column was the same.  This
does not guarantee that the same number of particles will be in each cell
of the grid, but the particle counts in cells will certainly vary less than
if we imposed a uniform grid, in which case the halo core would likely
occupy only a couple of cells, decreasing the efficacy of the partition.

For the shallower bound to the potential of particle $p$, we calculate the
potential produced by moving all other particles to the corners of their
cells farthest from $p$.  This potential is calculable in $O(n)$ time
because we only need to calculate the distance to a fixed (no dependence on
$n$) number of grid corners for each particle.  For the deeper bound, we
move all other particles to their cell corners closest to $p$, except for
particles in the same cell as $p$, whose contributions to the potential are
calculated directly.  For each particle, the number of these particles
requiring direct summation still scales with $n$, so in the worst case, the
deeper bound still takes $O(n^2)$ time.  However, the constant multiplying
$n^2$ is much less than in the direct calculation.  When deciding on the
number of grid partitions, there is a tradeoff between accuracy of the
upper and lower bounds and the time it takes to calculate them.  The number
of partitions should scale with the number of particles in the largest
halo, since more accuracy will be required to capture all of its structure.

In analysing a large cluster (see \S \ref{krav}), we found initially that
our unbinding algorithm destroyed a few haloes which were visually evident,
particularly when the full halo and not merely the core zone was used to
calculate the velocity centroid.  This happened because the haloes sent to
the unbinding algorithm contained high-velocity particles which were not
bound to the object, skewing the initial estimate of the velocity centroid
and thus unduly unbinding particles which were needed to keep the halo
together.

We addressed this problem by unbinding only the most unbound particles at
each iteration, a technique suggested by Kravtsov (private communication).
The only parameter-free way of doing this is to unbind only the most
unbound particle at every iteration.  We include code to do this with the
{\scshape voboz} package, even though it is cumbersome in practice since
the binding energy for each bound particle must be recalculated each time.
A time-saving alternative which we used is to set the threshold on
boundness so that at first, only extremely unbound particles escape, but
then to lower the threshold gradually to the right level.  To do this, we
find the most unbound particle, and set a multiplier $m$ equal to the ratio
of its kinetic to potential energy.  The role of $m$ is to inflate the
unbinding threshold artificially by multiplying the potential energy by it.
Before each iteration, we reduce $m$ by dividing it by a parameter $f>1$,
until $m=1$, and the true unbinding criterion appears.  The effects of
changing $f$ depend on the resolution and range of velocities in the
simulation, and will be discussed in an extreme case in \S \ref{krav}.

\section{Tests}

Using the NCSA p690 supercomputer, we applied {\scshape voboz} to a set of
nested 256$^3$-particle $\Lambda$CDM simulations (described in NHG), with
box sizes $32$, $64$, $128$, and 256 $h^{-1}$ Mpc.  We divided each
simulation into two parts in each dimension; calculating the particle
volumes and adjacencies took about one hour and 10 GB of RAM on each
octant; it thus took about 8 hours total, trivially split on to 8
processors.  The memory required can be reduced at the expense of processor
time if the simulation is split into more pieces.  The next step of the
analysis was to join the zones together, which took, in all cases, about 6
minutes, and 3.2 GB of RAM.  The time spent in the unbinding step depended
greatly on the size of the haloes in the simulation, i.e.\ on the mass
resolution and threshold density $\rho_{min}$ (lower threshold densities
give larger haloes).  In the 32 $h^{-1}$ Mpc simulation, using $f =
\sqrt{2}$, the unbinding step took from 4 to 65 processor-hours as
$\rho_{min}$ varied from 400 to 50; over the same range of $\rho_{min}$,
the 256 $h^{-1}$ Mpc simulation took only from 1 to 4 hours.  The amount of
processor time also depends on the value of the unbinding delicacy
parameter $f$.  For the results below, we used $f=\sqrt{2}$.  We did not
check that this choice unbinds particles with maximal delicacy (which we
did do for the cluster discussed in \S \ref{krav}), but we do not expect
the results of this simulation to saturate at values of the parameters as
extreme as those for the cluster in \S \ref{krav}, which has a higher
velocity dispersion.

\subsection{Mass Functions}\label{masstests}
\massfn
\massfnone
\mcomp
\rcdf

As Fig.\ \ref{massfn} shows, the mass spectra of {\scshape voboz} haloes
from these simulations, using a density cut-off of $\rho_{min} = 100$, is
roughly consistent with the Sheth, Mo \& Tormen (2001) analytical mass
function.  While the mass function from the 32 $h^{-1}$ Mpc simulation fits
almost exactly, there is a systematic increase in the number of haloes of a
given physical mass with box size.  This also occurs, to a slightly lesser
degree, in the {\scshape denmax} mass functions, indicating that it could
arise from decreasing mass resolution in the simulations, and perhaps not
from poor behavior by the HFA's.

The mass function does change with density cut-off $\rho_{min}$, but only
significantly changes for the largest haloes.  Figure \ref{massfnone} shows
how the mass function of {\scshape VOBOZ} haloes in the 32 $h^{-1}$ Mpc
simulation changes as $\rho_{min}$ varies from $400$ to $50$.  Generally,
the number of low-mass haloes stays fixed; the main effect of this
parameter is on the slope of the mass function.

It is also useful to see how {\scshape VOBOZ} mass varies with mass
resolution.  In NHG, we identified four haloes present in all of the nested
32, 64, 128, and 256 $h^{-1}$ Mpc simulations we ran.  Figure \ref{mcomp}
shows their masses (normalized to the particle mass in the 256 $h^{-1}$ Mpc
simulation) as a function of box size (and, therefore, mass resolution),
for four different values of $\rho_{min}$.  For all values of $\rho_{min}$,
{\scshape voboz} returned a halo mass less dependent on mass resolution
than did {\scshape denmax}.  For the canonical density cut-off, $\rho_{min}
= 100$ (the bold line in all panels), the mass of the halo changed little
with box size.

We had hoped that the unbinding criterion would take away the dependence on
$\rho_{min}$, and that as the density cut-off $\rho_{min}$ is decreased,
the bound mass of a large halo would plateau.  However, adding any
particles to a halo deepens its potential well, feeding back positively on
the number of particles possibly bound to it.  Thus, the bound mass of an
isolated halo usually rises unchecked with the pre-unbinding mass.  The
unbinding test is still quite necessary, however, to eliminate haloes whose
high velocity dispersions render them evanescent.  From the
$16{,}777{,}216$ particles in the 32 $h^{-1}$ Mpc simulation, $876{,}592$
peak particles, and thus zones, were detected; on average, one in every 19
particles was a peak.  The unbinding step reduced the number of haloes to
about $280{,}000$ ($\pm$ 800 as $\rho_{min}$ varied from $50$ to $400$).
In the 256 $h^{-1}$ Mpc simulation, $900{,}139$ zones were detected, out of
which $618{,}100 \pm 100$ were bound.  It is not surprising that more bound
structures were detected in the simulation encompassing the larger physical
volume.  Figure \ref{rcdf} shows the cumulative distribution of the ratio
$r$ of peak density to strongest link density for all haloes and all bound
haloes in the 32 $h^{-1}$ Mpc simulation, using $\rho_{min} = 100$.  At the
low-probability end, the CDF of pre-unbinding haloes (the dotted curve)
tracks the CDF from the Poisson simulation, but the CDF of bound haloes
(the dashed curve) departs from the Poisson curve, indicating that many of
the least probable haloes are not bound.  This supports our claim that $r$
is a good tracer of halo probability.

\subsection{Correlation Functions}

Because there is no resolution limit in {\scshape voboz}, we expected to,
and did, find pairs of haloes much closer together than in the algorithm we
previously applied to this simulation, {\scshape denmax}$^2$ (see NHG).  In
{\scshape denmax}$^2$, {\scshape denmax} is run as usual, and then is run
again, with half the canonical smoothing length, on each halo from the
previous run to resolve sub-haloes.  Figure \ref{crls} shows the
correlation functions (CFs) of sets of haloes characterized by cut-offs in
probability (assigned using the ratio $r$), compared to the CF of all
{\scshape denmax}$^2$ haloes.  {\scshape denmax}$^2$ imposes a strict
10-particle minimum; the {\scshape denmax}$^2$ haloes are larger and
therefore have higher CFs.  We also measured the total CF of haloes,
weighting pairs of haloes by the product of their probabilities; this total
CF is close to the unweighted CF of all (without a
probability cut-off) haloes.  We should note that these CFs are lower in
amplitude than are observed galaxy CFs, which stems from the inclusion of
even the smallest, two-particle haloes in the simulation, which have a mass
of $5 \times 10^8 M_\odot$.  The population of {\scshape
denmax}$^2$ haloes from this simulation which best fit the PSC$z$ power
spectrum (NHG) had at least 800 particles.

\crls
\haloterm

Figure \ref{crls} shows another trend as we vary the probability cut-off,
written in terms of the standard Gaussian $\sigma$.  Starting from a high
probability cut-off and decreasing it to allow in more dubious haloes, a
`hook' rises in the CF at low radius.  This says that the closest pairs
include at least one low-probability halo, which makes sense by our
definition of probability: a low-probability peak has a shallow, and most
likely short, landbridge to another, denser peak.  At the same time, the CF
decreases at high separation when improbable haloes are added, because they
are also generally smaller, and therefore more weakly clustered.
Interestingly, the radius separating these two regimes is relatively
constant for plausibly low values of $\sigma$, producing a nexus (here, at
about $0.1 h^{-1}$ Mpc) where CFs cross.  It is tempting to interpret the
`hook' at low radius as a `1-halo' or `Poisson' term (e.g.\ Zehavi et al.\
2004) in the galaxy CF, consisting of pairs of galaxies (i.e.\ subhaloes)
within the same halo.

To test this hypothesis, we tried to use a probability cut-off to produce a
`2-halo' term obtained explicitly using the halo model formalism from an
$N$-body simulation.  Kravtsov et al.\ found the 2-halo contribution to the
CF of haloes (and subhaloes) exceeding $10^{12}$ $h^{-1} M_\odot$ in an 80
$h^{-1}$ Mpc simulation, which appears in their Fig.\ 8.  In our Fig.\
\ref{haloterm}, we show CFs of all haloes from our 64, 128, and 256
$h^{-1}$ Mpc simulations (black curves), along with attempts at 2-halo
terms (grey curves).  To obtain the 2-halo term, we varied a probability
cut-off in increments of $\sigma$ to get the CF to turn away from the full
CF at about 1 $h^{-1}$ Mpc, where Kravtsov et al.\ found that their 1-halo
and 2-halo terms cross.  The probability cut-offs for the CFs shown are
$4\sigma$, $6\sigma$, and $7\sigma$ for the 256, 128, and 64 $h^{-1}$ Mpc
simulations, respectively.  This supports the idea that halo probability is
a measure of a halo's `sub-haloness,' although the cut-off at which full
`haloness' occurs varies with mass resolution, and is likely somewhat
fuzzy.  Another thing to point out is that the scale of the inflection
indicating the onset of the CF 1-halo term increases with the size of the
halo sample; as Figs.\ \ref{crls} and \ref{haloterm} show, the scale of the
inflection increases by a factor of ten as the mass cut-off is raised from
5 $\times 10^8 M_\odot$ to $10^{12}$ $h^{-1} M_\odot$.

We should also note that in Fig.\ \ref{haloterm}, the CFs of haloes
exceeding the same physical mass from three simulations of different box
size and mass resolution coincide over a wide range of scales, a
concordance which we could not achieve in NHG using {\scshape denmax} mass.
This boosts our confidence in {\scshape voboz}'s mass estimate.

\subsection{A Large Cluster}\label{krav}

We have claimed that {\scshape voboz} is adept at finding small structures
in simulations; to test this claim, we have applied it to a large,
high-resolution cluster of mass a few times $10^{14}$ $h^{-1} M_\odot$,
provided by Andrey Kravtsov.  It was drawn from a simulation appearing in
Tasitsiomi et al.\ (2004), which has a box size of 80 $h^{-1}$ Mpc, and a
particle mass ranging from $3.159 \times 10^8$ $h^{-1} M_\odot$ to 64 times
that.  The simulation was designed so that the smallest particles would end
up in clusters such as this one.  We ran {\scshape voboz} only on these
smallest particles, which comprised over 99\% of the mass at all radii out
to 2 $h^{-1}$ Mpc from the cluster core.  In the following analysis, we
considered only haloes within this radius.

In this region with a high velocity dispersion, it was necessary to unbind
particles delicately, unbinding only the most unbound particles at every
iteration.  As described at the end of \S \ref{unbinding}, we did this by
gradually decreasing a factor multiplying the potential energy until the
true unbinding criterion is left.  Figure \ref{ubtest} shows how the number
of haloes detected depends on the unbinding coarseness parameter $f$ (the
factor by which the potential multiplier is divided at each iteration).
Indeed, it saturates when $f$ is sufficiently small, which probably arises
from particle discreteness, since only one particle can be unbound at a
time.  We recommend using a reasonably small value of $f$, but it does not
matter much if $f$ moderately exceeds the saturation point, since most of
the haloes missed with a large choice of $f$ have low probability.  One
might also wonder how the masses of robust haloes changes with $f$; with a
couple of exceptions, they change not at all, or only negligibly.  For the
results discussed below, we used $f = \sqrt[128]{2}$.  A simulation without
as high a velocity dispersion would likely accomodate a larger value of
$f$.  We used an extremely low density cut-off $\rho_{min} = 1$, so that
the only halo which would be affected by $\rho_{min}$ would be the cluster
itself.

\ubtest
\kravmassfn
\kravmscat

Figure \ref{kravmassfn} shows the mass function of haloes in this cluster
returned by both {\scshape voboz} and a variant (Kravtsov et al.\ 2004) of
{\scshape bdm}.  The {\scshape bdm} halo list was produced by Kravtsov,
analysing only the smallest particles in the cluster as we did.  The mass
functions from both algorithms agree reassuringly well, except at the
low-mass end, where {\scshape voboz} detects far more haloes (if we apply
no probability cut-off).  However, these are also the most improbable
haloes, as illustrated by the descent of the solid curves with increasing
probability cut-off.  We also tried to match haloes detected by both
algorithms in a rather crude fashion: for each {\scshape bdm} halo, we
formed a list of all {\scshape voboz} haloes within 0.02 $h^{-1}$ Mpc of
it, which is a bit under twice the separation between the cluster core as
detected by {\scshape voboz} and the core as detected by {\scshape bdm}.
We declared the matching {\scshape voboz} halo to be the one with the
nearest mass on a logarithmic scale within this sphere.  This method is
`rather crude' because one {\scshape voboz} halo can match several
{\scshape bdm} haloes.  Out of 383 {\scshape bdm} haloes, 5 small ones did
not have {\scshape voboz} neighbors within the search radius.  Figure
\ref{kravmscat} shows a scatter plot of the masses of matching haloes,
which is again reassuring.

\weirdhalo

Our method of joining together zones is designed for the idealized
situation of Fig.\ \ref{tank}.  However, when it is applied to a collection
of middling haloes with a background density above the cut-off, the halo
which happens to have the highest-density core particle can acquire
unwelcome neighbors.  Figure \ref{weirdhalo} shows an extreme case of this
phenomenon, taken from the Kravtsov cluster.  All of the visually evident
subhaloes pass the $4\sigma$ test, but the probability we assign may start
to lose its meaning by $7\sigma$, since the most obvious subhalo does not
exceed this probability.  The `halo' pictured here has perhaps more
structure in velocity space than in position space, suggesting that an
ideal HFA would search for clusters in 6-D position-velocity space.  Such a
thing could separate two colliding haloes, which, in the worst case,
{\scshape voboz} would identify as a single halo, and then perhaps unbind
completely because of its bimodal velocity distribution.  In the mean time,
the unbinding method we use seems to work fairly well in picking out
structures in velocity space.

Figure \ref{weirdhalo} also illustrates the need to average together only
the velocities in the central zone of a halo to find the velocity centroid
used in unbinding.  With the velocity centroid defined in this way, the
same set of particles was returned as bound for all values of the unbinding
delicacy parameter $f$.  On the other hand, using the entire halo to
determine the velocity centroid, we obtained results which depended
strongly on $f$.  Sometimes, we would obtain the right answer (the halo
around the central zone); sometimes, another halo would be returned
(resulting in double counting in the halo catalog); and sometimes, all
particles would be unbound.

\section{Conclusion}

We have developed a halo-finding algorithm, called {\scshape voboz}, which
is nearly parameter-free, and which has a resolution limited only by the
discreteness of particles in the simulation it is analysing.  The Voronoi
diagram allows us to fix the densities and the sets of neighbors for all
particles in a `natural,' parameter-independent way, on arguably the finest
possible scale that contains meaningful structure.  Further degrees of
freedom are eliminated by assigning to each halo (or subhalo) a probability
that it exists, i.e.\ that it did not arise from Poisson noise.

Of the two parameters in {\scshape voboz}, one of them exists only to save
processor time, and needs not be used if one has no processor time
constraints.  This parameter controls the delicacy with which particles are
unbound, and the results saturate at the `right' answer (the parameter-free
situation in which one particle is unbound at a time) when the unbinding
becomes sufficiently delicate.  Additionally, most of the extra haloes
uncovered at extreme delicacy have meager probabilities.  The remaining
parameter is a density cut-off, necessary because haloes do not extend to
low densities in the real universe.  However, it only affects the masses of
the largest haloes in the simulation, and not the masses or detection of
subhaloes, which is where reliability in halo-finding algorithms is most
needed.

An ideal halo-finding algorithm would find groups of particles in both
position and velocity space, or even consider neighboring time slices in a
simulation.  In {\scshape voboz}, the velocities are used only to decide if
particles are energetically bound to haloes found in position space.  This
is an approximate criterion, but it seems to work acceptably well.

The probability {\scshape voboz} returns for each halo is certainly tied to
the finite mass resolution of the simulation it is analysing.  However, as
we decrease a cut-off in halo probability, an interesting signal emerges in
the halo correlation function which resembles the 1-halo term in the halo
model of large-scale structure.  This suggests an interpretation of halo
probability as a measure of a halo's `sub-haloness,' which may increase
{\scshape voboz}'s appeal for researchers of the halo model.

The {\scshape voboz} code is publically available, at
\url{http://casa.colorado.edu/~neyrinck/voboz/}.

\section*{Acknowledgments}

We thank Andrey Kravtsov for enlightening discussions, and for providing us
with cluster data to analyze, without which we would not have realized how
delicately haloes must be unbound.  We also thank Rien van de Weygaert for
a helpful referee's report.  This work was supported by NASA ATP award
NAG5-10763, NSF grant AST-0205981, and a grant by the National
Computational Science Alliance.

\appendix
\section{Guarding sub-boxes}

\guard
\guardclose

To check that dividing the simulation box does not alter its Voronoi
diagram, we surround each sub-box with a buffer zone large enough to
contain all of the neighbors of the particles inside the sub-box.  Figure
\ref{guard} shows a sub-box surrounded by a buffer.  We have deployed guard
particles, shown as diamonds, inside the buffer.  If one of the guard
particles is returned as a neighbor to a sub-box particle, then it is
possible that the sub-box particle has a neighbor outside the buffer, and
we must recalculate the VD on the sub-box with a larger buffer.

The distance $g$ between the sub-box and the guard particles is determined
as follows.  The guard particles are arranged inside each face of the
buffer, of width $b$, on two dimensional grids with spacing $s$.  Figure
\ref{guardclose} shows two diagonally adjacent (and thus separated by a
distance $s\sqrt{2}$) guard particles, shown as diamonds.  We want the
guard particles to `catch' any sub-box particles for which a particle
outside the buffer could affect its Voronoi cell.  The worst-case scenario,
in which the guard particles have the least guarding power, occurs if there
is a particle right on the border of the sub-box, where the square is.  The
closest possible point to the square outside the buffer is the triangle.
If the triangle is in the square's set of neighbors (but is artificially
excluded because it lies outside the buffer), then so will a guard point if
the perpendicular bisector (a dotted line) between that guard point and the
square intersects the line segment between the triangle and square at its
midpoint (the confluence of the three lines of length $b/2$), or nearer to
the square than the triangle.  So the guard points must be placed on a
sphere with radius $b/2$, tangent to both the buffer and sub-box
boundaries.  This leads to an equation for the largest-possible $g$:
\begin{equation}
  g = \frac{b}{2}\left(1+\sqrt{1 - \frac{2s^2}{b^2}}\right).
  \label{geq}
\end{equation}
As the number of guard points increases, $s$ decreases, which moves $g$
toward the edge of the buffer.  Thus, increasing the number of guard points
can provide an alternative to increasing the buffer size if guard points
are encountered in the tessellation, but only if the buffer truly contains
all of the neighbors of points inside the sub-box.

\end{document}